\documentclass[journal]{IEEEtran}

\usepackage[usenames, dvipsnames]{color}
\usepackage{amsmath,amsfonts}
\usepackage[caption=false,font=normalsize,labelfont=sf,textfont=sf]{subfig}
\usepackage{textcomp}
\usepackage{stfloats}
\usepackage{verbatim}
\usepackage{array}
\usepackage{enumitem}
\usepackage{soul}
\usepackage{cite}
\usepackage{siunitx}
\usepackage{adjustbox}
\usepackage{hhline}
\usepackage{makecell}
\usepackage{caption}
\usepackage{float}
\usepackage{steinmetz}
\usepackage{tablefootnote}
\usepackage{scrextend}
\usepackage{multirow}
\usepackage{url, times, booktabs, tabularx, amssymb, bm, bbm}
\usepackage{algorithm, algorithmic}
\usepackage{bbding}
\usepackage{graphicx}
\usepackage{ragged2e}
\usepackage[colorlinks=true, linkcolor=blue, urlcolor=blue, citecolor=blue]{hyperref}

\def\BibTeX{{\rm B\kern-.05em{\sc i\kern-.025em b}\kern-.08em
    T\kern-.1667em\lower.7ex\hbox{E}\kern-.125emX}}
\usepackage{balance} 

\begin{document}

\title{PSELDNets: Pre-trained Neural Networks on a Large-scale Synthetic Dataset for Sound Event Localization and Detection}

\author{Jinbo Hu,~\IEEEmembership{Student Member,~IEEE}, Yin Cao,~\IEEEmembership{Member,~IEEE}, Ming Wu,~\IEEEmembership{Member,~IEEE}, \\ Fang Kang,~\IEEEmembership{Member,~IEEE}, Feiran Yang,~\IEEEmembership{Member,~IEEE}, Wenwu Wang,~\IEEEmembership{Senior Member,~IEEE} \\ Mark D. Plumbley,~\IEEEmembership{Fellow,~IEEE}, Jun Yang,~\IEEEmembership{Senior Member,~IEEE}
\thanks{
\scriptsize{

Jinbo Hu, Ming Wu, and Jun Yang are with the Key Laboratory of Noise and Vibration Research, Institute of Acoustics, Chinese Academy of Sciences, Beijing 100190, China (e-mail: hujinbo@mail.ioa.ac.cn; mingwu@mail.ioa.ac.cn; jyang@mail.ioa.ac.cn). Jinbo Hu and Jun Yang are also with the University of Chinese Academy of Sciences, Beijing 100049, China. \textit{(Corresponding author: Jun Yang, Yin Cao)}

Yin Cao is with the Department of Intelligent Science, Xi’an Jiaotong Liverpool University, Suzhou 215123, China (e-mail: yin.k.cao@gmail.com).

Fang Kang is with the Center for Machine Vision and Signal Analysis (CMVS), University of Oulu,  Oulu 90570, Finland (e-mail: fang.kang@oulu.fi).

Feiran Yang is with the State Key Laboratory of Acoustics, Institute of Acoustics, Chinese Academy of Sciences, Beijing 100190, China (feiran@mail.ioa.ac.cn).

Wenwu Wang and Mark D. Plumbley are with the Centre for Vision, Speech and Signal Processing, University of Surrey, Guildford GU2 7XH, U.K. (e-mail: w.wang@surrey.ac.uk, m.plumbley@surrey.ac.uk).

This work was supported by the National Key Research and Development Project (NO. 2022YFB2602003), Grant XJTLU RDF-22-01-084, and Engineering and Physical Sciences Research Council (EPSRC) Grant EP/T019751/1. 
}}}

\maketitle

\begin{abstract}

Sound event localization and detection (SELD) has seen substantial advancements through learning-based methods. These systems, typically trained from scratch on specific datasets, have shown considerable generalization capabilities. Recently, deep neural networks trained on large-scale datasets have achieved remarkable success in the sound event classification (SEC) field, prompting an open question of whether these advances can be extended to the development of SELD foundation models. In this paper, leveraging the power of pre-trained SEC models, we propose pre-trained SELD networks (PSELDNets) on a large-scale synthetic dataset. The synthetic dataset, generated by convolving sound events with simulated spatial room impulse responses (SRIRs), contains 1,167 hours of audio clips with an ontology of 170 sound classes. These PSELDNets are applied to various SELD scenarios. When we adapt PSELDNets to specific scenarios, particularly in cases of low-resource data, we introduce a data-efficient fine-tuning method, AdapterBit. PSELDNets are evaluated on \textit{synthetic-test-set} using collected SRIRs from the TAU Spatial Room Impulse Response Database (TAU-SRIR DB) and achieve satisfactory performance. We also carried out experiments to validate the transferability of PSELDNets to three publicly available datasets and our own real-world recordings. The results demonstrate that PSELDNets surpass state-of-the-art systems across all publicly available datasets. Given the need for direction-of-arrival estimation, SELD generally relies on sufficient multi-channel audio clips. However, incorporating the AdapterBit, PSELDNets show more efficient adaptability to various scenarios using minimal multi-channel or even just monophonic audio clips, outperforming traditional fine-tuning approaches.

\end{abstract}

\begin{IEEEkeywords}
Sound event localization and detection (SELD), pre-trained SELD networks, data-efficient fine-tuning.
\end{IEEEkeywords}

\section{Introduction}

Sound event localization and detection (SELD) combines sound event detection (SED) with direction-of-arrival (DOA) estimation, with the goal of recognizing the categories, onsets, offsets, and DOAs of various sound sources. SELD frameworks represent audio sources in both spatial and temporal domains, making them suitable for applications such as robot listening, audio surveillance, and smart home environments.

\subsection{Existing learning-based SELD methods}

In recent years, there have been notable advancements in learning-based SELD methods. Adavanne et al. \cite{Adavanne2018_JSTSP} introduced SELDnet, an end-to-end network designed for simultaneous sound event detection and DOA estimation. Nevertheless, SELDnet faces challenges in identifying overlapping sound events of the same class from different locations. To address this homogenous overlap issue, the Event-Independent Network V2 (EINV2) is proposed \cite{cao2020event, cao2021, hu2022track}. EINV2 uses a track-wise output format and permutation invariant training to predict a single sound event and its corresponding DOA for each track. Unlike SELDnet and EINV2, the Activity-coupled Cartesian DOA (ACCDOA) combines SED and DOA tasks into a single output and embeds activity information into Cartesian DOA vectors \cite{shimada2021accdoa}. The Multi-ACCDOA (mACCDOA) \cite{multiaccdoa} extends ACCDOA by incorporating a track-wise output format and employing auxiliary duplicated permutation invariant training to address the homogenous overlap problem.

On the other hand, numerous learning-based SELD investigations \cite{Cao2019, cao2020event, cao2021, nguyen2021salsa, shimada2021accdoa, multiaccdoa, hu2022track, hu22dw, wang2023four, wang2023dcase} have predominantly utilized synthetic datasets from SELD challenge events \cite{dcase2019task3, dcase2020task3, dcase2021task3, l3das21, l3das22, l3das23}, showing promising performance in both simulated and real spatial environments. However, these systems have two limitations. Firstly, the target sound event classes that the systems predict must be pre-specified before training, posing a challenge since each application scenario may require different target classes. Secondly, learning-based SELD approaches can suffer from performance degradation when exposed to acoustic environments not encountered during training, i.e., a phenomenon known as environment shift \cite{hu2023selective}.

One of the effective ways to address the problems of unknown sound event classes and unseen acoustic environments is by acquiring significant scenario-specific data for training. However, creating spatial sound event signals is a complex task involving extensive data collection and computational generation. This process requires convolving dry sound source signals with measured spatial room impulse responses (SRIRs). Moreover, manually collecting and annotating real-world spatial sound event recordings is very costly, and publicly accessible real-scene SELD data is limited \cite{starss22, starss23}. To mitigate these challenges, the zero-and-few-shot SELD system \cite{zsl-seld} and environment-adaptive Meta-SELD \cite{meta-seld, hu2023selective} utilize pre-trained models to operate effectively with limited data. Despite these developments, there is still a notable lack of foundation models for SELD. In contrast, several foundation models have recently been developed \cite{kong2020panns, ast, passt, htsat} for sound event classification (SEC), which focus on identifying the categories of sound events without concerning their timestamps (i.e., onset and offset). These SEC models are highly relevant to SELD tasks, but their potential benefits for SELD systems have not been adequately investigated.

\subsection{Foundation models in SEC}
Deep neural networks have made substantial strides in SEC research \cite{kong2020panns, ast, passt, htsat}. A key milestone was the introduction of AudioSet \cite{gemmeke2017audio}, which is a comprehensive dataset featuring over 2 million human-annotated 10-second audio clips and an ontology of 527 sound classes, and utilized for general-purpose sound event recognition. Convolutional neural networks (CNNs), exemplified by Pre-trained Audio Neural Networks (PANNs) \cite{kong2020panns}, extract local features from audio spectrograms and enhance performance by optimizing the network's depth and breadth.

Recently, Transformer architectures \cite{transformer}, which have proven effective in sequence modeling, have been adapted to computer vision by partitioning images into smaller patches \cite{vit, swinTransformer}. Inspired by these approaches, several studies, such as the Audio Spectrogram Transformer (AST) \cite{ast}, the Patchout faSt Spectrogram Transformer (PaSST) \cite{passt}, and the Hierarchical Token-Semantic Audio Transformer (HTS-AT) \cite{htsat}, apply purely attention-based models to audio spectrograms to capture long-range global context. AST \cite{ast} leverages the self-attention mechanism, overlapping patches from audio spectrograms, and pre-trained parameters from computer vision to build the first convolution-free model for SEC. Drawing inspiration from SpecAugment \cite{specaug} and the mask technique used in Bidirectional Encoder Representations from Transformers (BERT) \cite{bert}, PaSST \cite{passt} offers an efficient implementation of AST by omitting segments of the Transformer's input sequence during training. This method encourages the Transformer to classify the events using an incomplete sequence. In comparison, HTS-AT \cite{htsat} uses Swin Transformer blocks with shifted window attention \cite{swinTransformer}, enhancing efficiency by limiting self-attention calculations to local non-overlapping windows while allowing cross-window connections. These models achieve state-of-the-art (SOTA) SEC results on AudioSet. 

Furthermore, these models, which were pre-trained on large-scale datasets, offer the potential for transferability to other audio-related tasks to further improve performance \cite{kong2020panns, ast, passt, htsat}, such as acoustic scene classification, music genre classification, and speech emotion classification. Nevertheless, the efficient transfer of these pre-trained models to various audio tasks remains challenging. One common method for adapting the pre-trained models to these tasks involves fine-tuning all the parameters of the pre-trained models using the datasets designed for these tasks. However, this technique requires significant computational resources and memory capacity. On the other hand, it can result in a loss of model generalization, possibly due to catastrophic interference among tasks \cite{adaptformer}.

\subsection{Parameter-efficient fine-tuning}

To mitigate the challenges associated with efficient transfer, the parameter-efficient fine-tuning (PEFT) methodology, which only fine-tunes a small number of (extra) parameters to achieve strong performance, has been extensively investigated across the domains of natural language processing \cite{lora, unified_view_peft, bitfit, adapter_nlp} and computer vision \cite{aim, adaptformer, 1vs100, vpt}. 

Prominent PEFT methods include Low-Rank Adaptation (LoRA) \cite{lora}, Adapter tuning \cite{aim, adaptformer, adapter_nlp}, prompt tuning \cite{vpt}, and others. The fundamental principle of these PEFT methodologies involves freezing the primary or all pre-trained parameters and introducing additional trainable parameters for fine-tuning. Expanding on these PEFT techniques, various researchers working on audio processing have integrated some model-specific Adapters into their frameworks \cite{AAT, adapter_conformer, adapter_asv, comparison_PEFT_google}. The Adapter, a straightforward plug-and-play module, is designed for attention-based networks and entails incorporating a few lightweight bottleneck networks into the Transformer layers. These methodologies retain the generality of the pre-trained model, conserve computational resources, reduce data requirements, and attain competitive or even superior performance.

\subsection{Our contributions}

In this study, we endeavor to develop SELD foundation models that can be applied to various real-world scenarios. We introduce pre-trained SELD networks (PSELDNets) on a large-scale synthetic dataset. This dataset, comprising approximately 1,167 hours of audio clips and featuring an ontology of 170 sound classes, is scenario-agnostic and generated by convolving sound event clips from FSD50K \cite{fonseca2021fsd50k} with simulated SRIRs. The PSELDNets, inheriting the architectures of our previously proposed CNN-Conformer \cite{hu2022track, hu22dw} and pre-trained models that achieve SOTA results in SEC, such as PANNs \cite{kong2020panns}, PaSST \cite{passt} and HTS-AT \cite{htsat}, extract spatial and global features from multi-channel spectrograms. Unlike common model architectures tailored for the SELD problem, e.g., CRNN \cite{nguyen2021salsa} and CNN-attention \cite{wang2023four, hu2022track}, to our knowledge, PSELDNets represent the first SELD model that exploits the Transformer model. We evaluate the performance of PSELDNets on \textit{synthetic-test-set} that uses measured SRIRs from TAU Spatial Room Impulse Response Database (TAU-SRIR DB) \cite{srir-db} and obtain satisfactory performance.

We transfer PSELDNets to multiple publicly available datasets, including the Detection and Classification of Acoustic Scenes and Events (DCASE) 2021 Challenge Task 3 \cite{dcase2021task3}, the L3DAS22 Challenge Task 2 \cite{l3das22}, the Sony-TAu Realistic Spatial Soundscapes 2023 (STARSS23) \cite{starss23} dataset, and our own audio recordings. Experimental results demonstrate the transferability of PSELDNets, showing that the models consistently exceed SOTA benchmarks \cite{dcase2021results, shimada2021ensemble, dcase2023results, wang2023four, hu2022track} across all these publicly available datasets.

Inspired by PEFT techniques \cite{aim, adaptformer, bitfit}, we introduce a data-efficient fine-tuning method for SELD, AdapterBit, which fine-tunes only the additionally inserted multi-layer perceptron (MLP) Adapter and bias terms, enabling the efficient utilization of low-resource data, including minimal multi-channel or even monophonic clips. By employing AdapterBit for transfer learning in specific SELD scenarios using low-resource data, PSELDNets exhibit superior performance, compared to the conventional full fine-tuning methods. Notably, when utilizing monophonic clips, pseudo-multi-channel clips are generated by convolving the audio sources with the theoretical responses of the microphone array to ensure compatibility with the input to PSELDNets.

The contribution of this work includes:
\begin{enumerate}
    \item Synthesizing a large-scale SELD dataset designed to include numerous sound event instances and various acoustic environments.
    \item Introducing PSELDNets trained on the large-scale synthetic SELD dataset to develop foundation models.
    \item Transferring PSELDNets to several SELD scenarios and achieving SOTA performance.
    \item Proposing a data-efficient fine-tuning technique to adapt PSELDNets to specific scenarios using limited data.
    \item Releasing the source code, the pre-trained parameters of PSELDNets, and the large-scale synthetic SELD dataset\footnote{\url{https://github.com/Jinbo-Hu/PSELDNets}}.
\end{enumerate}

\section{Data synthesis}
The synthesis of SELD clips is achieved through the convolution of clean sound event clips from FSD50K \cite{fonseca2021fsd50k} with simulated SRIRs. The important components for accurately simulating spatial sound event recordings are the acquisition of high-quality sound event clips and SRIRs.

\subsection{Sound event clips}
\label{sec: selection_criteria}
Various efforts have been dedicated to developing audio datasets for SEC \cite{FSDnoise18k, FSDKaggle2019, fonseca2021fsd50k, gemmeke2017audio, ESC_50, SONYC_UST_V2, nigens, chime_home}. In this study, we select sound event clips based on the AudioSet Ontology \cite{gemmeke2017audio} and emphasize strong labeling, single-source clips and high label quality.

\subsubsection{AudioSet Ontology}
We focus on creating SELD foundation models. To meet this objective, the selected classes need to cover a comprehensive range of everyday sounds and be scalable regarding data and vocabulary. Accordingly, we use the AudioSet Ontology\footnote{\url{https://research.google.com/audioset/ontology/index.html}} for organizing our data. The AudioSet Ontology includes 632 sound event classes, arranged hierarchically with up to 6 levels, and encompasses a variety of everyday sounds. The class annotations in the datasets, e.g., AudioSet \cite{gemmeke2017audio}, FSDnoisy18K \cite{FSDnoise18k}, FSDKaggle2019 \cite{FSDKaggle2019}, and FSD50K \cite{fonseca2021fsd50k}, make use of the vocabularies provided by this ontology.

\subsubsection{Strong data labeling}
The SELD task, akin to the SED task, necessitates predicting the exact start and end times of sound events, which is essential for accurately predicting the trajectory of moving sound sources. Nonetheless, audio datasets that provide such detailed timestamp annotations \cite{nigens, audioset_strong}, i.e., strong labels, are quite rare. Most datasets provide annotations at the clip level without precise timestamps, i.e., weak labels \cite{gemmeke2017audio, fonseca2021fsd50k, FSDnoise18k, FSDKaggle2019, SONYC_UST_V2, ESC_50, chime_home}.

\subsubsection{Single-source clips}
\label{sec: synth_sample_ss}
Individual sound sources are spatially isolated to be distinguished in SELD. Conversely, typical audio datasets usually include audio clips annotated with multiple class labels, indicating that each clip may contain several overlapping sound events or a single event with hierarchical-propagation labels \cite{fonseca2021fsd50k, gemmeke2017audio}. When synthesizing SELD clips, each selected sound event clip must contain only a single sound source at a time to guarantee an accurate representation of spatialized sound events.

\subsubsection{Label quality}

The quality of datasets is important for model performance. Early studies on audio datasets often relied on small and exhaustively annotated datasets \cite{nigens, ESC_50, chime_home}, and as large-scale datasets like AudioSet \cite{gemmeke2017audio} have emerged, label inaccuracies have become more common \cite{fonseca2021fsd50k}, due to the impracticality of exhaustive manual annotation. Some datasets \cite{FSDnoise18k, FSDKaggle2019} focus on learning under noisy label conditions, which is out of the scope of this work. Label accuracy remains critical for ensuring model reliability, as noisy labels can introduce interference and lead to performance degradation \cite{label_noise}. This work prioritizes label quality while maintaining a substantial dataset size.

Based on the above considerations, we select single-source clips from FSD50K \cite{fonseca2021fsd50k} for synthesis. FSD50K encompasses a collection of 51,197 audio clips totaling 108 hours, manually labeled with 200 classes derived from the AudioSet Ontology. Despite its weak labeling, FSD50K exhibits a high label density \cite{shah2018closerlookweaklabel-key}. Label density refers to the portion of the recording duration during which the annotated sound event is actually present. The clips with high label density allow us to treat sound events throughout the entire clip as active. Notably, strong annotations in AudioSet \cite{audioset_strong} are not used for data synthesis, due to significant imbalance and incompleteness in certain classes.

\subsection{Spatial room impulse responses}

SELD generally necessitates multi-channel audio inputs for effective source localization. First-order ambisonics (FOA), well known as an array-agnostic format, is widely employed in various SELD datasets \cite{dcase2019task3, dcase2020task3, dcase2021task3, starss22, starss23, l3das21, l3das22, l3das23, wearable_seld_datasets}. Numerous advanced methods utilize FOA signals, instead of original microphone-array-format signals, to achieve SOTA results \cite{wang2023four, shimada2021ensemble, wang2023dcase, wang2024dcase}. Additionally, several studies \cite{amb_encoding_arb, neural_amb_encoding} explore ambisonics encoding of arbitrary microphone arrays, yielding competitive performance. Therefore, we employ FOA-format SRIRs to synthesize SELD clips.

Ambisonics represents a data format that decomposes a sound field on an orthogonal basis of spherical harmonic functions. This format is typically derived by converting the spherical microphone array signals \cite{rafaely2015fundamentals}. The FOA signal comprises four channels $(W, Y, Z, X)$ with $W$ denoting an omnidirectional microphone and $(Y, Z, X)$ referring to three bidirectional microphones aligned along the Cartesian axes. The theoretical spatial responses of FOA are \cite{rafaely2015fundamentals, archontisPhD, dcase2019task3}:
\begin{equation}
\begin{aligned}
& H_1(\phi, \theta)=1, \\
& H_2(\phi, \theta)=\sin (\phi) \cos (\theta), \\
& H_3(\phi, \theta)=\sin (\theta), \\
& H_4(\phi, \theta)=\cos (\phi) \cos (\theta),
\end{aligned}
\label{eq: FOA_response}
\end{equation}
where $\theta$ and $\phi$ denote elevation and azimuth. 

The computational generation method for FOA-format SRIRs involves a two-step process: microphone-array RIRs simulation and ambisonics format conversion. Microphone-array RIRs are generated using the image source method \cite{allen1979image}. The procedure for converting microphone-array signals to FOA signals can be found in \cite{rafaely2015fundamentals, archontisPhD, politis2017comparing, koyama2022spatial, hu2023selective}.

\section{The SELD systems}
\subsection{Related SELD methods}
\label{sec: seld_output_format}

We introduce two existing learning-based SELD methodologies: Event-Independent Network V2 (EINV2) \cite{cao2020event, cao2021, hu2022track} and Activity-coupled Cartesian DOA (ACCDOA) \cite{shimada2021accdoa, multiaccdoa}. Both approaches perform frame-wise prediction of the temporal activity and spatial trajectory of sound events. For simplicity, subsequent illustrations omit the temporal dimension and instead focus on the SELD output format for individual frames.

\subsubsection{EINV2}

\begin{figure}
    \centering
    \includegraphics[width=0.98\linewidth]{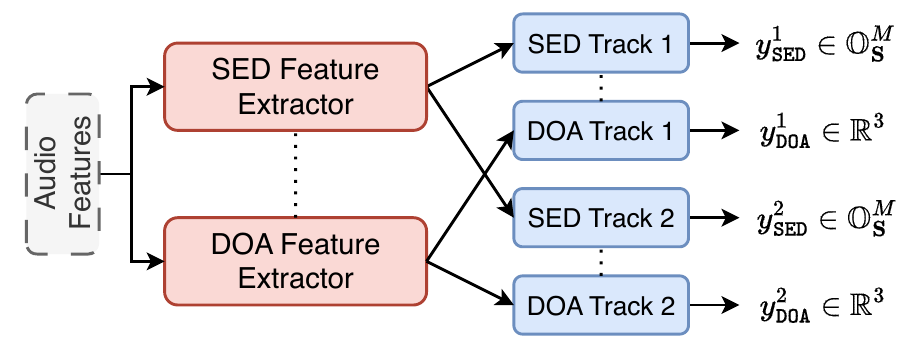}
    \caption{The architecture of EINV2. The solid line boxes represent trainable neural networks. The dotted lines are learnable parameters connecting two branches.}
    \label{fig: EINV2}
\end{figure}

\begin{figure}
    \centering
    \includegraphics[width=0.98\linewidth]{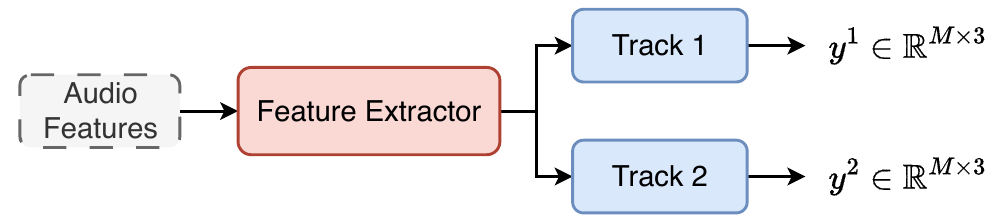}
    \caption{The mACCDOA representation of the SELD model. There is no track dimension in the ACCDOA representation.}
    \label{fig: mACCDOA}
\end{figure}

EINV2 \cite{cao2021} comprises two branches: SED and DOA estimation, which are linked through a soft parameter-sharing strategy, e.g., multiple sets of learnable parameters. Each branch has multiple event-independent tracks, forming track pairs. Each pair can only predict a sound event with its corresponding DOA. Permutation-invariant training is used to handle the track misalignment between the ground truth and the prediction. The architecture is shown in Fig. \ref{fig: EINV2}. For the $i$-th track, $y_{\mathtt{SED}}^i$ indicates one-hot encoding of $M$ sound event classes in the set $\mathbf{S}$, and $y_{\mathtt{DOA}}^i$ represents Cartesian DOA output. The number of tracks depends on the maximum number of overlapping events.

\subsubsection{ACCDOA}

The ACCDOA approach represents the presence of a sound event through the amplitude of its corresponding Cartesian DOA vector. Unlike EINV2, the ACCDOA representation merges two branches into one, thereby obviating the necessity of balancing the loss between SED and DOA branches and avoiding an increase in model parameters. 

However, the ACCDOA representation cannot detect multiple instances of the same event type from various locations. To address this issue, the mACCDOA representation has been proposed \cite{multiaccdoa}. mACCDOA integrates both class-wise and track-wise output formats, as shown in Fig. \ref{fig: mACCDOA}. Additionally, auxiliary duplicating permutation invariant training (ADPIT) is introduced to tackle problems of track misalignment and sparse target outputs.

\subsection{Network architectures}

The SEC field has seen substantial advancements due to deep neural networks \cite{kong2020panns, ast, passt, htsat} and large-scale datasets \cite{gemmeke2017audio, fonseca2021fsd50k}. Utilizing pre-trained models that exhibit superior performance in SEC, e.g., PANNs \cite{kong2020panns}, AST \cite{ast}, PaSST \cite{passt} and HTS-AT \cite{htsat}, may improve the SELD task. Consequently, the structures of PSELDNets align with these pre-trained SEC models for effective transfer learning. PSELDNets take as input the concatenation of log-mel spectrograms and intensity vectors extracted from FOA signals, and predict active sound events with corresponding DOA vectors for each timestamp, adhering to SELD output formats described in Sec. \ref{sec: seld_output_format}.

\subsubsection{PANNs}
PANNs \cite{kong2020panns} are convolution-based models trained from scratch on AudioSet. Current SELD techniques predominantly utilize CNN-attention hybrid models, which have demonstrated superior performance, e.g., ResNet-Conformer \cite{wang2023dcase, wang2023four} and our previously proposed CNN-Conformer \cite{hu2022track, hu22dw}. 

Following the above model architectures, we extend CNN-Conformer to CNN14-Conformer, where a Conformer block \cite{gulati2020conformer} follows the main body of the CNN14 \cite{kong2020panns} module. CNN14 contains a stack of 6 VGG-like \cite{vgg} CNN blocks, and the Conformer comprises two feed-forward layers with residual connections sandwiching the multi-head self-attention and convolution modules. The CNN block extracts local fine-grained features, while the Conformer block captures both local and global context dependencies in an audio sequence. 

\subsubsection{PaSST}

\begin{figure}[t]
    \centering
    \centerline{\includegraphics[width=0.98\linewidth]{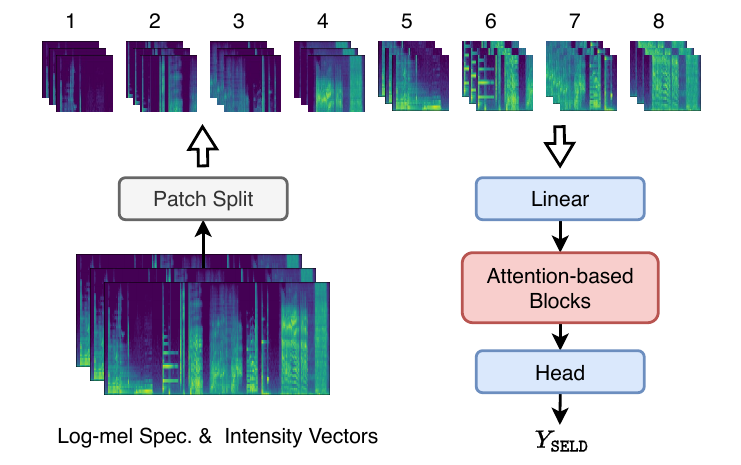}}
    \caption{The architecture of the purely attention-based network.}
    \label{fig: atte_backbone}
\end{figure}

PaSST \cite{passt} is an advanced and efficient variant of AST \cite{ast}. The overall architecture of PaSST is shown in Fig. \ref{fig: atte_backbone}, and the specific structure of the attention-based blocks is detailed in Fig. \ref{fig: attention}(a). The 2D audio spectrogram is split into a sequence of overlapping patches, which are subsequently flattened and linearly projected to a sequence of 1D patch embeddings. These embeddings are processed using a standard Transformer Encoder \cite{transformer}. Additionally, in analogy to the mask technique used in BERT \cite{bert} and SpecAugment \cite{specaug}, PaSST adopts unstructured Patchout and structured Patchout, where unstructured Patchout randomly omits patches from any position, and structured Patchout picks some frequency bins or time frames and removes the corresponding row or column of extracted patches. These two approaches improve generalization and reduce computation and memory complexity.

We employ PaSST with the structured Patchout approach \cite{passt} in the frequency axis to ensure the valid output of each time frame. PaSST utilizes the classification and distillation token mechanism for label prediction, inherently preventing it from predicting event start and end times in audio clips. The output embeddings from the Transformer layer contain temporal information for each frame. To tackle the prediction of sound events with precise timestamps, we project the final layer embeddings into SELD output formats via a linear head.

\subsubsection{HTS-AT}
\begin{figure}[t]
    \centering
    \centerline{\includegraphics[width=0.95\linewidth]{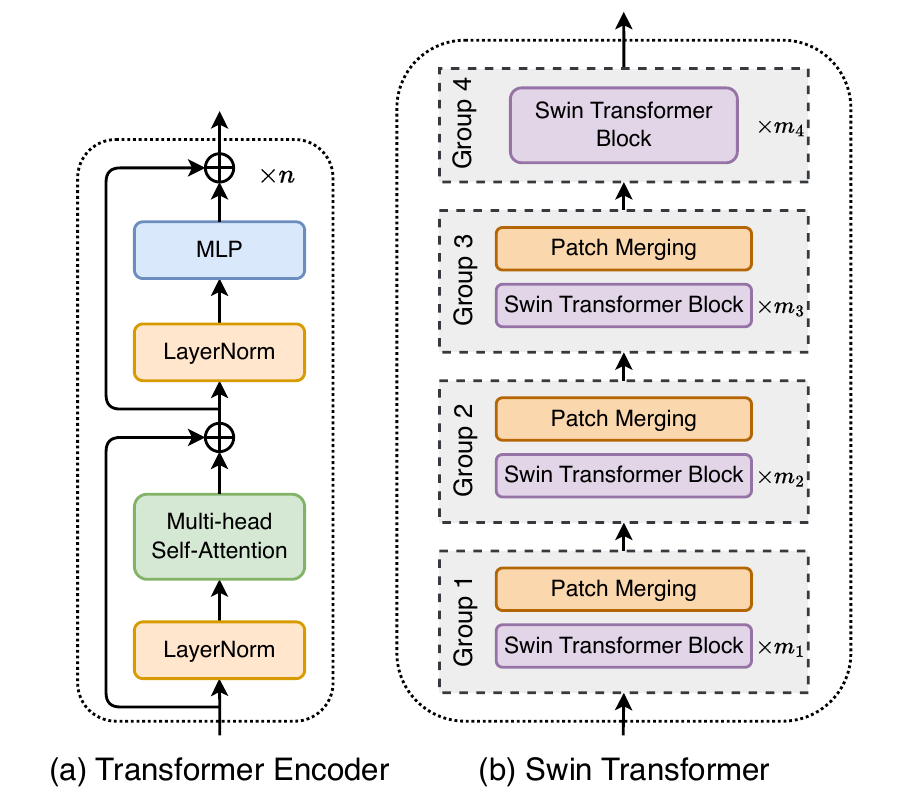}}
    \caption{The detailed architecture of the attention-based blocks in Fig. \ref{fig: atte_backbone}. The Transformer Encoder is employed in AST and PaSST, while the Swin Transformer is employed in HTS-AT.}
    \label{fig: attention}
\end{figure}

HTS-AT \cite{htsat} combines the Swin Transformer \cite{swinTransformer} and a token-semantic module \cite{tscam}. The Swin Transformer focuses on self-attention within each local window, which comprises fixed-size and non-overlapping patches. A key design element of the Swin Transformer is the shifted window attention across successive self-attention layers, introducing connections between neighboring non-overlapping windows from the preceding layer. Moreover, Swin Transformer builds hierarchical feature maps by gradually merging neighboring patches in deeper Transformer layers to reduce the sequence size. The token-semantic module \cite{tscam} employs a simple convolution layer as the head layer and converts output feature maps from the final Swin Transformer Block into activation maps for the prediction of each timestamp. The details of HTS-AT are depicted in Fig. \ref{fig: atte_backbone} and Fig. \ref{fig: attention}(b), with the detailed architecture of the Swin Transformer Block being identical to the standard Transformer Encoder in Fig. \ref{fig: attention}(a). 

\subsection{Data augmentation}
Data augmentation is a valuable technique for improving the generalization capabilities of a system. Given the successful application of our previously proposed data augmentation chains \cite{hu2022track, hu22dw} in L3DAS22 Task 2 \cite{l3das22} and DCASE 2022 Task 3 \cite{starss22}, we adopt this technique to increase the data diversity.

Each data augmentation chain comprises various augmentation operations, which are randomly selected and linked together. Following the methodology described in \cite{hu2022track, hu22dw}, we randomly sample $k=3$ augmentation chains and select Mixup \cite{zhang2018mixup}, Cutout \cite{zhong2020random}, SpecAugment \cite{specaug}, and frequency shifting \cite{nguyen2021salsa} as data augmentation operations. Additionally, we employ the rotation of FOA signals \cite{mazzon2019first, wang2023four} as an independent spatial augmentation operation, which is not a part of data augmentation chains. 

\section{Data-efficient Fine-tuning}

Fine-tuning involves deploying a pre-trained model for a new task, where all parameters are initialized from the pre-trained model, with the possible exception of the head layer. The conventional full fine-tuning approach may result in the loss of model generalization, potentially due to catastrophic interference among tasks \cite{adaptformer}. Inspired by PEFT techniques \cite{adaptformer, aim, bitfit, adapter_nlp}, we introduce a data-efficient fine-tuning strategy, AdapterBit. 

SELD generally necessitates multi-channel audio inputs for source localization. By utilizing AdapterBit, PSELDNets can be more efficiently adapted to various SELD scenarios using limited data, with a particular emphasis on the monophonic sound event clips. Specifically, when employing monophonic signals for fine-tuning, we generate pseudo-FOA signals based on theoretical responses of the employed microphone array to align with the input to PSELDNets.

\begin{figure}[t]
    \centering
    \centerline{\includegraphics[width=0.95\linewidth]{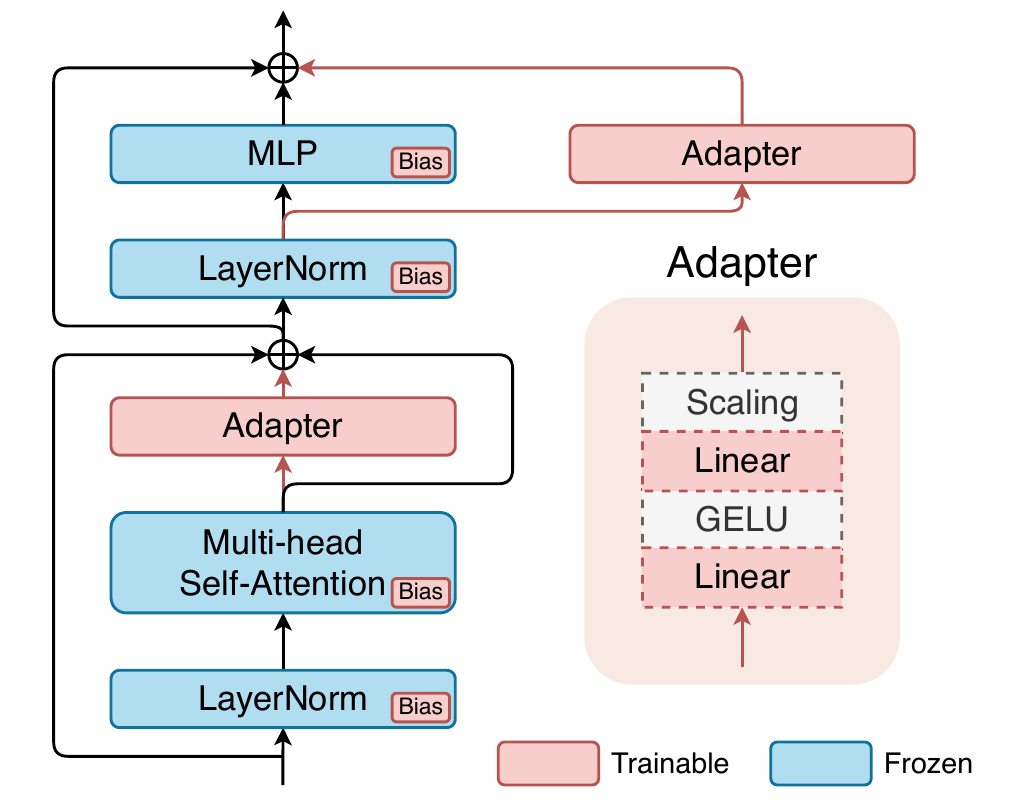}}
    \caption{The brief illustration of AdapterBit.}
    \label{fig: AdapterBit}
\end{figure}

\subsection{AdapterBit}
We design the AdapterBit structure as depicted in Fig. \ref{fig: AdapterBit}. AdapterBit integrates MLP Adapters following the architecture outlined in \cite{AAT, adaptformer, aim}, while additionally introducing bias-term tuning. In bias-term tuning, only the bias terms from the pre-trained model, such as those in linear and convolution layers, are fine-tuned. The Adapter is constructed with a Gaussian Error Linear Unit (GELU) non-linearity positioned between two linear layers, which is subsequently integrated into the standard Transformer Encoder layer through a scaling factor $s$. The scaling factor serves to balance the general features produced by the frozen branch and the scenario-specific features produced by the trainable branch. For a given input feature $x_l$, the Adapter produces adapted features $\tilde{x_l}$ as follows:
\begin{equation}
    \tilde{x_l} = s \cdot (W_2 \cdot \mathtt{GELU}(W_1 \cdot x_l + b_1) + b_2),
\end{equation}
where $W_1$, $W_2$, $b_1$ and $b_2$ denote the weights and biases of the linear layers. The parameters $W_1$ and $b_1$ are randomly initialized, while $W_2$ and $b_2$ are set to zero. The rationale behind zero initialization is to ensure that the adapted model keeps the same state as the pre-trained model at the beginning of fine-tuning by utilizing the parallel design and the residual connection of the Adapter.

During the fine-tuning stage, we focus on optimizing just the newly added parameters and the bias term from PSELDNets, while keeping the other parameters frozen. Specifically, PSELDNets initializes its weights from the pre-trained checkpoint and maintains all parameters, except the bias terms, in a frozen state. The Adapters and the bias term of the adapted model are updated using the SELD data from the specific scenario, such as different microphone array types, acoustic environments and sound event categories. During the inference phase, we reload all pre-trained parameters, including those that were previously kept frozen, along with the newly inserted and fine-tuned parameters.

\subsection{Pseudo-FOA signals}

SELD generally necessitates multi-channel audio input to enable effective DOA estimation, e.g., four-channel FOA signals input to PSELDNets. To fulfil this input requirement, we can generate pseudo-FOA signals from monophonic signals by utilizing the theoretical responses of FOA described in Eq. (\ref{eq: FOA_response}). Additionally, the monophonic signals must contain non-overlapping sound events to satisfy the single-source clip requirement described in Sec. \ref{sec: synth_sample_ss}.

The pseudo-FOA signals are obtained as follows:
\begin{equation}
\left[\begin{array}{l}
W(t, f) \\
Y(t, f) \\
Z(t, f) \\
X(t, f)
\end{array}\right]=\left[\begin{array}{c}
1 \\
\sin (\phi) \cos (\theta) \\
\sin (\theta) \\
\cos (\phi) \cos (\theta)
\end{array}\right] S(t, f),
\end{equation}
where $S(t, f)$ denotes the monophonic signal spectrogram. $\phi$ and $\theta$ can be randomly sampled from the desired distribution of the azimuth and elevation. 

The pseudo-FOA signals can be considered a form of regularization for input monophonic signals, as they preserve information related to sound events while mitigating the loss of inter-channel connections wherever possible.

\section{Experimental Setups}

\subsection{Datasets}
\label{sec: datasets}

Audio clips from FSD50K \cite{fonseca2021fsd50k} are selected according to the criteria in Sec. \ref{sec: selection_criteria}. We select single-source sound event clips and filter out the classes with fewer than 30 clips and those that pose recognition challenges. As a result, a total of 170 classes are selected. The selected audio clips comprise 31,444 samples for training, amounting to 43.4 hours, and an additional 3,701 samples for testing, totaling 5.3 hours. 

SRIRs are mainly generated via simulation \cite{pyroomacoustics}. We simulate diverse shoebox-shaped rooms employing frequency-dependent absorption coefficients. This approach avoids the requirement for sampling from a distribution of reverberation times and estimating absorption coefficient values, thereby preventing unrealistic scenarios such as long reverberation times in small rooms. Absorption materials from typical acoustic material databases\footnote{\url{https://www.acoustic-supplies.com/absorption-coefficient-chart/}}$^,$\footnote{\url{https://pyroomacoustics.readthedocs.io/en/pypi-release/pyroomacoustics.materials.database.html}} are randomly allocated to the wall, ceiling and floor surfaces of each simulated room. We use these simulated SRIRs to synthesize spatialized static sound sources for training. To generate overlapping data, we convolve each monophonic single-source clip with SRIRs from different spatial locations within the same room configuration and then mix the resulting signals. Furthermore, additional spatialized sound events, including moving sources, are synthesized using collected SRIRs from TAU-SRIR DB \cite{srir-db} to evaluate the simulated SRIRs and various network architectures. We adopt the publicly available code for data synthesis\footnote{\url{https://github.com/Jinbo-Hu/SELD-Data-Generator}}.

In total, we synthesize 67,000 1-minute clips amounting to approximately 1,117 hours for training, where each clip is simulated using a unique room configuration, termed as \textit{synthetic-training-set}. Additionally, we synthesize 3,060 1-minute clips amounting to roughly 51 hours for testing, denoted as \textit{synthetic-test-set}. The distribution of maximum polyphony of 1, 2, and 3 in the synthetic dataset follows a ratio of approximately 10:5:2.

\subsection{Hyper-parameters}
The sampling rate is 24 kHz. We extract 64-dimension log mel spectrograms from FOA signals using a Hanning window of 1024 points and a hop size of 240. Each audio clip is segmented to a fixed duration of ten seconds for training and inference. All hyper-parameters of PSELDNets are consistent with those in the pre-trained SEC models \cite{kong2020panns, passt, htsat}. The number of event-independent tracks is 3 in EINV2 and mACCDOA. A batch size of 32 and the AdamW \cite{adamw} optimizer are employed for training. The learning rate is set to $10^{-4}$ for the first 20 epochs and subsequently decreased to $10^{-5}$ for the following 5 epochs.

\subsection{Evaluation metrics}
\label{sec: metric}
We use a joint metric of localization and detection\cite{mesaros2019joint, overviewofDCASE} in this work: two location-dependent detection metrics, F-score ($\mathrm{F}_{ 20^\circ}$) and error rate ($\mathrm{ER}_{ 20^\circ}$), and two class-dependent localization metrics, localization recall ($\mathrm{LR}_\mathrm{CD}$) and localization error ($\mathrm{LE}_\mathrm{CD}$). In contrast to the standard computation of F-score and error rate in SED \cite{SEDMetric}, $\mathrm{F}_{ 20^\circ}$ and $\mathrm{ER}_{ 20^\circ}$ consider a prediction a true positive only if the sound event is accurately detected and its estimated direction lies within a spatial threshold of $20^\circ$ away from the ground truth, while the predictions falling outside this threshold are treated as false positives. $\mathrm{LE}_\mathrm{CD}$ and $\mathrm{LR}_\mathrm{CD}$ compute the mean angular error and true positive rate in the case when sound event classes are predicted correctly. Note that $\mathrm{LR}_\mathrm{CD}$ can also be interpreted as the unthresholded recall. 

We use an aggregated SELD metric for the method comparison and hyper-parameter selection: 
\begin{equation}
\scalebox{0.95}{$
{\mathcal{E}_\mathtt{SELD}}=\frac{1}{4}\left[\mathrm{ER}_{ 20^{\circ}}+\left(1-\mathrm{F}_{ 20^{\circ}}\right)+\frac{\mathrm{LE}_{\mathrm{CD}}}{180^{\circ}}+\left(1-\mathrm{LR}_{\mathrm{CD}}\right)\right].$}
\end{equation}

For comparison and consistency across different task setups, a macro-average of $\mathrm{F}_{ 20^\circ}$, $\mathrm{LR}_\mathrm{CD}$, $\mathrm{LE}_\mathrm{CD}$, and $\mathcal{E}_\mathtt{SELD}$ across classes is utilized in STARSS23 and \textit{synthetic-test-set}, and a micro-average of those metrics across instances is employed in other datasets. An effective system should demonstrate a low $\mathrm{ER}_{ 20^{\circ}}$, a high $\mathrm{F}_{ 20^{\circ}}$, a low $\mathrm{LE}_{\mathrm{CD}}$, a high $\mathrm{LR}_{\mathrm{CD}}$, and a low ${\mathcal{E}_\mathtt{SELD}}$. 
 
\section{Experiments}

\begin{table*}[t]
    \centering
    \caption{Description of the utilized SELD models involving various stages.}
    \label{tab:model_desc}
    \setlength{\arrayrulewidth}{0.6pt}
    \begin{tabular}{|c|c|c|}
        \hline
         Method & Initial parameters & Datasets \\
         \hline
         \multirow{1}{*}{Pre-train} & Pre-trained SEC checkpoints & \multirow{1}{*}{\textit{synthetic-training-set} (1,117 h)}\\ 
         \hline
         From-scratch & Pre-trained SEC checkpoints & \multirow{3}{*}{\makecell[c]{One dataset from \{\textit{L3DAS22 Task 2} (7.5 h), \textit{DCASE 2021 Task 3} (13.3 h), \textit{STARSS23} (7.5 h),\\ scenario-specific synthetic datasets (1 to 45 h)\} for different experiments}}\\
         \cline{1-2}
         \multirow{1}{*}{\makecell[c]{Fine-tune}} & \multirow{2}{*}{\makecell[c]{Pre-trained SELD checkpoints}} & \\
         \cline{1-1}
         AdapterBit & & \\
         \hline
    \end{tabular}
\end{table*}

Firstly, the performance of PSELDNets is evaluated on \textit{synthetic-test-set}, investigating various networks and SELD output formats. Secondly, PSELDNets are transferred to multiple publicly available datasets. Subsequently, the efficiency of the data-efficient fine-tuning approach is validated on low-resource data. Finally, the effectiveness of PSELDNets and the data-efficient fine-tuning approach are tested using our own audio recordings, termed \textit{Indoor Recordings}. 

The SELD models involving various stages are presented in Table \ref{tab:model_desc}. If not specified, the Pre-train method in this work denotes the SELD models trained on \textit{synthetic-training-set}, initialized with pre-trained SEC checkpoints, such as those from PANNs, PaSST or HTS-AT. The From-scratch, Fine-tune, and AdapterBit methods represent different strategies for applying these SELD models to specific SELD scenarios. The From-scratch method utilizes pre-trained SEC checkpoints, specifically HTS-AT in this work, without using any pre-trained SELD checkpoints. In contrast, Fine-tune and AdapterBit methods directly utilize the pre-trained SELD checkpoints, specifically those integrating HTS-AT with the mACCDOA output format in this work.

\subsection{Results on the synthetic dataset}

\begin{table}[t]
    \centering
    \caption{Results of various networks with the mACCDOA representations.}
    \begin{adjustbox}{width=\columnwidth,center}
        \begin{tabular}{c|c|cccc|c}
            \toprule[1pt]
            Network& \# Params& $\mathrm{ER}_{20^{\circ}}\downarrow$ & $\mathrm{F}_{20^{\circ}}\uparrow$ &
            $\mathrm{LE}_\mathrm{CD}\downarrow$ & $\mathrm{LR}_\mathrm{CD}\uparrow$ & $\mathcal{E}_{\mathtt{SELD}}\downarrow$\\
            \midrule 
             CNN14-Conformer& 179.4 M& 0.805& 25.4\%& $\mathbf{17.3^\circ}$& 32.1\%&  0.582\\
             PaSST& 52.3 M& \textbf{0.773}& \textbf{29.2\%}& $17.6^\circ$& 33.2\%& \textbf{0.562} \\
             HTS-AT& 34.6 M& 0.784& 27.6\%& $17.6^\circ$& \textbf{33.9}\%& 0.567  \\
             \bottomrule[1pt]
        \end{tabular}
    \end{adjustbox}
    \label{tab: backbone}
\end{table}

\begin{table}[t]
    \centering
    \caption{Results of HTS-AT with various SELD methods.}
    \begin{adjustbox}{width=\columnwidth,center}
        \begin{tabular}{c|cccc|c}
            \toprule[1pt]
            Method & $\mathrm{ER}_{20^{\circ}}\downarrow$ & $\mathrm{F}_{20^{\circ}}\uparrow$ &
            $\mathrm{LE}_\mathrm{CD}\downarrow$ & $\mathrm{LR}_\mathrm{CD}\uparrow$ & $\mathcal{E}_{\mathtt{SELD}}\downarrow$\\
            \midrule 
             ACCDOA& \textbf{0.777}& \textbf{27.9}\%& $17.1^\circ$& 33.0\%& \textbf{0.566}  \\
             mACCDOA& 0.784& 27.6\%& $17.6^\circ$& \textbf{33.9}\%& 0.567 \\
             EINV2& 0.801& 24.7\%& $\mathbf{15.4^\circ}$& 25.3\%&  0.597\\
             \bottomrule[1pt]
        \end{tabular}
    \end{adjustbox}
    \label{tab: seld_output_format}
\end{table}

\subsubsection{Network architecture}
We evaluate the performance of CNN14-Conformer, PaSST and HTS-AT, and choose the mACCDOA \cite{multiaccdoa} representations as the main SELD output formats. Each network employs its respective pre-trained SEC checkpoints \cite{kong2020panns, passt, htsat}, excluding the additional Conformer \cite{gulati2020conformer} module, which is randomly initialized due to the absence of an appropriate pre-trained checkpoint. 

Table \ref{tab: backbone} presents the comparison of various networks. We observe that PSELDNets achieve $\mathrm{LR}_\mathrm{CD}$ of over $32\%$ and $\mathrm{LE}_\mathrm{CD}$ of approximately $17^\circ$ on a macro-average across all classes. Notably, when comparing the result on \textit{synthetic-test-set} with those on scenario-specific datasets presented in Tables \ref{tab: L3DAS22}, \ref{tab: DCASE2021} and \ref{tab: STARSS23}, the performance gap in localization can be attributed to the discrepancy of SRIRs and source motion characteristics between \textit{synthetic-training-set} and \textit{synthetic-test-set}, as illustrated in Sec. \ref{sec: datasets}, whereas the performance gap in detection is primarily due to the diversity of sound events, spanning a total of 170 classes, as well as label noise, including imprecise timestamps and incorrect sound event annotations in FSD50K.

Although CNN14-Conformer has a substantial number of parameters, it performs the worst among the three networks, possibly due to the challenges in optimizing such a large model. Compared to PaSST, HTS-AT achieves similar performance but with fewer parameters. Consequently, we select HTS-AT as the baseline model for further investigation.

\subsubsection{SELD methods}

We evaluate the performance of three SELD methods employing HTS-AT: ACCDOA representations \cite{shimada2021accdoa}, mACCDOA representations \cite{multiaccdoa}, and EINV2 \cite{cao2021}. Within the EINV2 method, we utilize two relatively independent SED and DOA branches, both with identical architectures and pre-trained checkpoints. These branches are connected through several sets of trainable parameters \cite{cao2021} following each group of Swin Transformers \cite{swinTransformer} as illustrated in Fig. \ref{fig: attention}(b).

The results of various SELD methods are presented in Table \ref{tab: seld_output_format}. Among the three methods, the EINV2 method exhibits the worst performance, especially in detection. The EINV2 method has considerably worse $\mathrm{LR}_\mathrm{CD}$ but significantly better $\mathrm{LE}_\mathrm{CD}$ compared to the ACCDOA-based methods. One potential explanation for this discrepancy is that EINV2 utilizes multiple tracks, with each track predicting only one of 170 event classes, resulting in sparse outputs in the SED branch. In contrast, the mACCDOA representation is trained using auxiliary duplicated permutation invariant training (ADPIT), enabling each track to learn with the same target as the ACCDOA format. This training mechanism leads to dense outputs for mACCDOA, thereby achieving performance similar to that of ACCDOA.

\subsection{Transfer to various scenarios}
\label{sec: res_transfer_learning}
In this section, we investigate one application of PSELDNets. We employ HTS-AT with mACCDOA for transfer learning and apply it to several SELD scenarios using the full fine-tuning method. Notably, some systems only report their results using either ensemble or single models with post-processing. For a fair comparison, we adopt a post-processing method containing moving average (MA) and dynamic threshold (DT). During inference, test samples are segmented into 10-second clips with a 0.5-second hop length. The results for each 0.5-second segment are averaged across all corresponding time-overlapped segments, referred to as the MA method. Unlike the common approach that uses a uniform threshold for predicting sound event classes, DT employs class-specific thresholds.

For each scenario, we evaluate the following strategies: 1) Fine-tuning a model initialized from the SEC checkpoints pre-trained on AudioSet, denoted as the \textit{From-scratch} method; 2) Fine-tuning a model initialized from the SELD checkpoints pre-trained on \textit{synthetic-training-set}, referred to as the \textit{Fine-tune} method. The difference between the \textit{From-scratch} and \textit{Fine-tune} methods is illustrated in Table \ref{tab:model_desc}. Notably, the performance difference between these two methods highlights the improvements achieved through the pre-trained SELD models, as shown in the following results.

\subsubsection{L3DAS22 Task 2}

\begin{table}[t]
    \centering
    \caption{Results on L3DAS22 Task 2.}
    \begin{adjustbox}{width=\columnwidth,center}
    \begin{tabular}{c|c|cccc|c}
        \toprule[1pt]
         Method& Aug. &$\mathrm{ER}_{20^{\circ}}\downarrow$ & $\mathrm{F}_{20^{\circ}}\uparrow$ &
         $\mathrm{LE}_\mathrm{CD}\downarrow$ & $\mathrm{LR}_\mathrm{CD}\uparrow$ & $\mathcal{E}_{\mathtt{SELD}}\downarrow$\\
         \midrule
         \makecell[l]{(\#1) Hu et al.\cite{hu2022track}} &\CheckmarkBold &0.437 &65.1\% &$11.9^\circ$ &73.2\% &0.280  \\
         \midrule
         \multirow{2}{*}{From-scratch} &\XSolidBrush &0.617 &49.6\% &$17.9^\circ$ &67.6\% &0.386  \\
          &\CheckmarkBold &0.445 &63.8\% &$13.4^\circ$ &72.4\% &0.289  \\
         \midrule
         \multirow{2}{*}{Fine-tune} &\XSolidBrush &0.370 &70.6\% &$11.6^\circ$ &79.0\% &0.235  \\
          &\CheckmarkBold &0.361 &70.3\% &$12.2^\circ$ &77.0\% &0.239  \\
         \midrule
        \ \ Fine-tune $\ast$ &\CheckmarkBold & \textbf{0.330}& \textbf{73.6\%} &$\mathbf{11.3^\circ}$& \textbf{80.4\%}& \textbf{0.213}\\
        \bottomrule[1pt]
    \end{tabular}
    \end{adjustbox}
    \begin{justify}
        $\ast$ denotes post-processing methods. 
    \end{justify}
    \label{tab: L3DAS22}
\end{table}

L3DAS22 Task 2 \cite{l3das22} focuses on investigating 3D sound event localization and detection using two groups of FOA microphones in a large office room. The dataset, synthesized using measured SRIRs, contains 900 30-second audio recordings with 14 classes of sound events. Table \ref{tab: L3DAS22} shows the performance comparison between PSELDNets and our previously proposed system \cite{hu2022track}, CNN-Conformer, on the evaluation set. The latter system obtained the top rank in L3DAS22 Task 2. We modify the output format of CNN-Conformer to predict the corresponding DOAs rather than 3D Cartesian coordinates of corresponding sound events. For a fair comparison, we utilize only the centre FOA microphones and disregard the secondary FOA microphones. Experimental results indicate the From-scratch method performs comparably to CNN-Conformer \cite{hu2022track}, while the Fine-tune method surpasses CNN-Conformer by a large margin. Nonetheless, the Fine-tune method exhibits no performance improvement with data augmentation, possibly due to the high degree of similarity between the simulated environments in \textit{synthetic-training-set} and recording environments in the target dataset. Additionally, the post-processing method further improves performance. The Fine-tune method with post-processing achieves superior performance across all metrics.

\subsubsection{DCASE 2021 Task 3}
The dataset in DCASE 2021 Task 3 \cite{dcase2021task3}, synthesized using measured SRIRs, comprises 800 1-minute audio recordings. Different from the dataset in L3DAS22 Task 2, the DCASE 2021 Task 3 dataset encompasses moving sources and directional interference outside of the target classes. Table \ref{tab: DCASE2021} presents the performance difference between PSELDNets and the top two systems \cite{shimada2021ensemble, nguyen2021dcase, dcase2021results} on the evaluation set. Notably, these top two systems report exclusively the results of the ensemble models with post-processing. Experimental results reveal that both fine-tuned PSELDNets and the data augmentation method significantly improve performance. Moreover, when comparing the aggregated SELD metric $\mathcal{E}_\mathtt{SELD}$, our single model fine-tuned PSELDNets with post-processing perform even better than the SOTA system proposed by Shimada et al. \cite{shimada2021ensemble}, which was achieved by ensemble models with post-processing.

\begin{table}[t]
    \centering
    \caption{Results on DCASE 2021 Task 3.}
    \begin{adjustbox}{width=\columnwidth,center}
    \begin{tabular}{c|c|cccc|c}
        \toprule[1pt]
         Method& Aug. &$\mathrm{ER}_{20^{\circ}}\downarrow$ & $\mathrm{F}_{20^{\circ}}\uparrow$ &
         $\mathrm{LE}_\mathrm{CD}\downarrow$ & $\mathrm{LR}_\mathrm{CD}\uparrow$ & $\mathcal{E}_{\mathtt{SELD}}\downarrow$\\
         \midrule
         \makecell[l]{(\#1) Shimada et al. \cite{shimada2021ensemble,dcase2021results} $\ast\star$} &\CheckmarkBold &0.320 & \textbf{79.1\%} &$\mathbf{8.5^\circ}$& \textbf{82.8\%} &0.187  \\
         \midrule
         \makecell[l]{(\#2) Nguyen et al. \cite{nguyen2021dcase,dcase2021results} $\ast\star$} &\CheckmarkBold &0.320 &78.3\% &$10.0^\circ$ &78.3\% &0.202  \\
         \midrule
         \multirow{2}{*}{From-scratch} &\XSolidBrush &0.484 &60.5\% &$15.8^\circ$ &71.1\% &0.314  \\
          &\CheckmarkBold &0.435 &61.7\% &$17.1^\circ$ &80.3\% &0.278  \\
         \midrule
         \multirow{2}{*}{Fine-tune} &\XSolidBrush &0.394 &69.3\% &$12.9^\circ$ &76.7\% &0.251  \\
          &\CheckmarkBold &0.329 &74.7\% &$11.6^\circ$ &79.6\% &0.213  \\
        \midrule
        \ \ Fine-tune $\ast$ &\CheckmarkBold &\textbf{0.285} &79.0\% &$10.0^\circ$ &82.0\% &\textbf{0.183} \\
        \bottomrule[1pt]
    \end{tabular}
    \end{adjustbox}
    \begin{justify}
        $\ast$ denotes post-processing methods. $\star$ denotes ensemble models.
    \end{justify}
    \label{tab: DCASE2021}
\end{table}

\subsubsection{STARSS23}

\begin{table}[t]
    \centering
    \caption{Results on the STARSS23 dataset.}
    \begin{adjustbox}{width=\columnwidth,center}
    \begin{tabular}{c|cc|cccc|c}
        \toprule[1pt]
         Method& Ext. Data & Aug. &$\mathrm{ER}_{20^{\circ}}\downarrow$ & $\mathrm{F}_{20^{\circ}}\uparrow$ &
         $\mathrm{LE}_\mathrm{CD}\downarrow$ & $\mathrm{LR}_\mathrm{CD}\uparrow$ & $\mathcal{E}_{\mathtt{SELD}}\downarrow$\\
         \midrule
         \makecell[l]{(\#1) Wang et al. \cite{wang2023dcase,dcase2023results} $\ast$} &Wang-set &\CheckmarkBold &0.400 &	\textbf{64.0\%} &$\mathbf{13.4^\circ}$& 74.0\% &0.274  \\
         \midrule
         \makecell[l]{(\#2) Xue et al. \cite{xue2023dcase,dcase2023results} $\ast$} & Xue-set &\CheckmarkBold &0.430 &54.8\% &$14.7^\circ$ &68.0\% &0.321  \\
         \midrule
         \multirow{2}{*}{From-scratch} & - &\XSolidBrush &0.749 &24.1\% &$30.8^\circ$ &51.0\% &0.542  \\
          & - &\CheckmarkBold &0.530 &42.9\% &$18.2^\circ$ &58.6\% &0.404  \\
         \midrule
         \multirow{4}{*}{Fine-tune} & - &\XSolidBrush &0.640 &41.4\% &$20.7^\circ$ &62.7\% &0.429  \\
          & - &\CheckmarkBold &0.523 &54.4\% &$15.4^\circ$ &65.6\% &0.352  \\
          & Base-set &\CheckmarkBold &0.429 &56.5\% &$14.5^\circ$ &69.8\% &0.312  \\
          & + Synth-set &\CheckmarkBold &0.411 &58.2\% &$14.7^\circ$ &73.2\% &0.295  \\
          \midrule
          \ \ Fine-tune $\ast$ &\makecell[c]{Base-set \\ + Synth-set} &\CheckmarkBold &\textbf{0.390} &62.4\% &$14.4^\circ$ &\textbf{77.7\%} &\textbf{0.267} \\
         \bottomrule[1pt]
    \end{tabular}
    \end{adjustbox}
    \begin{justify}
        $\ast$ denotes post-processing methods.
    \end{justify}
    \label{tab: STARSS23}
\end{table}

The STARSS23 \cite{starss23} dataset, an extended version of the STARSS22 \cite{starss22} dataset, was collected in real-world environments, annotated manually, and served as the dataset for DCASE 2023 Task 3 and DCASE 2024 Task 3. Various clips in STARSS23 were recorded with combinations of moving and static participants acting in simple scenes and interacting among themselves and with the sound props. STARSS23 comprises roughly 7.5 hours of recordings in its development set. Due to the limited size of STARSS23, DCASE provides an additional official synthetic dataset \cite{official} for baseline training of Task 3 of DCASE 2022-2023. The official synthetic dataset, denoted as \textit{Base-set}, is simulated by convolving sound event clips with measured SRIRs from TAU-SRIR DB. Since the labels of the STARSS23 evaluation set are not publicly available, Table \ref{tab: STARSS23} shows results of the top two systems \cite{wang2023dcase, xue2023dcase, dcase2023results} and PSELDNets on the STARSS23 validation set. These top two systems report the results for the single models with post-processing and also synthesize substantial scenario-specific datasets for training, referred to as \textit{Wang-set} and \textit{Xue-set}. For a fair comparison, we also synthesize scenario-specific datasets, termed as \textit{Synth-set}. The SOTA system proposed by Wang et al. \cite{wang2023dcase}, incorporated a class-dependent sound separation approach \cite{ss_seld} into the SELD system. The Fine-tune method with post-processing outperforms the system of Wang et al. \cite{wang2023dcase} in terms of $\mathcal{E}_\mathtt{SELD}$, more specifically, in terms of $\mathrm{LR}_\mathrm{CD}$ and $\mathrm{ER}_{20^{\circ}}$.

\subsubsection{Discussions}

We investigate the performance of PSELDNets applied to these scenario-specific downstream datasets to evaluate the effects of data augmentation and post-processing. Additionally, we discuss the limitations of PSELDNets. 

\begin{figure*}[t]
    \centering
    \centerline{\includegraphics[width=\textwidth]{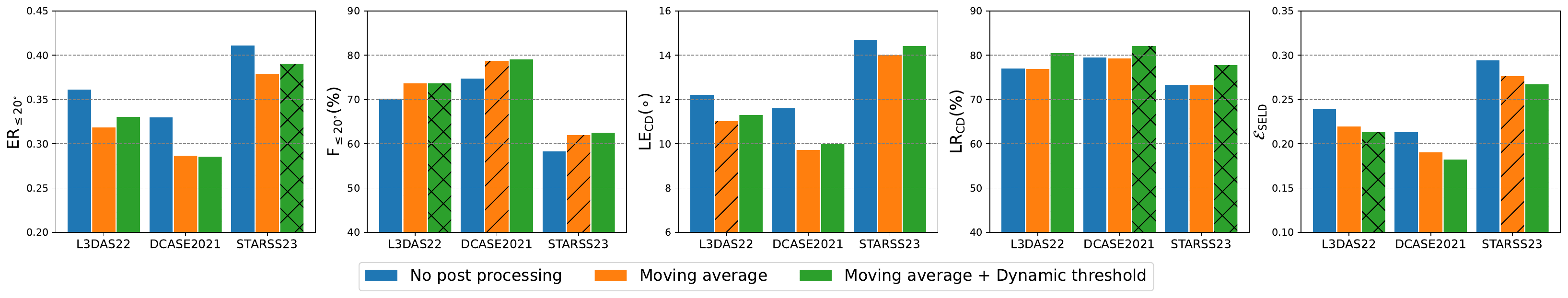}}
    \caption{The effect of the post-processing method on PSELDNets.}
    \label{fig: post_processing}
\end{figure*}

\begin{figure*}[t]
    \centering
    \centerline{\includegraphics[width=\textwidth]{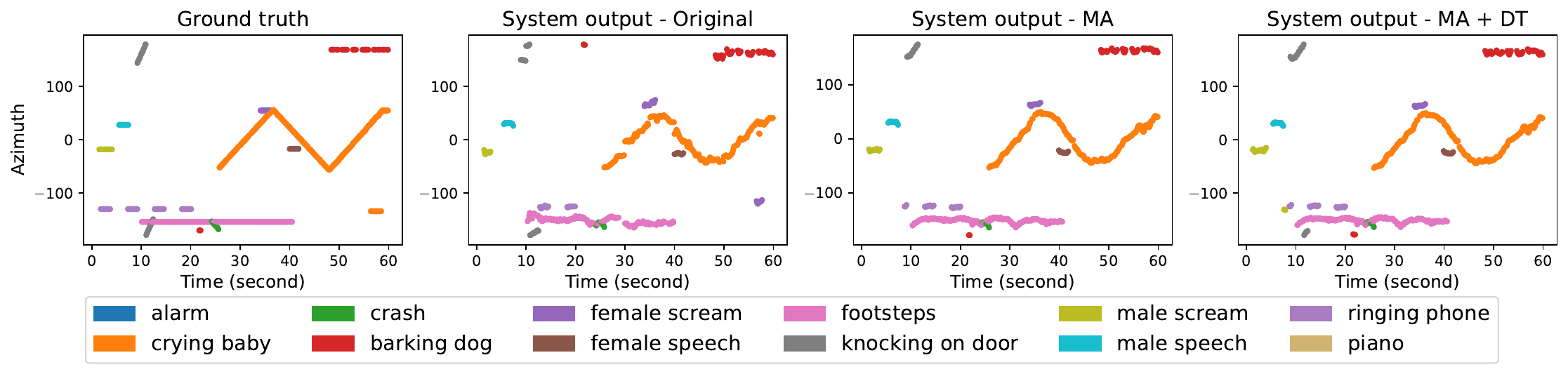}}
    \caption{Visualization of the ground truth and the system output w/ or w/o post-processing for a clip from the DCASE 2021 Task 3 evaluation set. SED predictions with the corresponding azimuth estimations are presented.}
    \label{fig: seld_outputs}
\end{figure*}

\textbf{Impact of data augmentation.} 
Empirical evidence indicates that learning-based SELD methods highly rely on large amounts of data, and data augmentation can increase the diversity of samples \cite{wang2023four, hu2022track}. Despite utilizing pre-trained checkpoints of PSELDNets, data augmentation continues to improve performance significantly, as shown in Tables \ref{tab: DCASE2021} and \ref{tab: STARSS23}. When applied to downstream datasets, PSELDNets offer general prior knowledge, while data augmentation provides a technique to effectively exploit scenario-specific data.

\textbf{Impact of post-processing.}
We surprisingly observe that the post-processing method can also provide significant performance improvement. Fig. \ref{fig: post_processing} presents the effect of the post-processing method in the above three datasets. We see that MA improves performance notably in $\mathrm{ER}_{20^{\circ}}$, $\mathrm{F}_{20^{\circ}}$ and $\mathrm{LE}_\mathrm{CD}$, while DT offers substantial performance improvement in $\mathrm{LR}_\mathrm{CD}$. Fig. \ref{fig: seld_outputs} illustrates a visualization of the ground truth and the system output for a clip from the evaluation set of DCASE 2021 Task 3. We present SED predictions along with the corresponding azimuth estimations. Overall, MA smooths the predicted trajectories of sound events, bringing them closer to the ground truth, compared to the model output without post-processing, such as the moving trajectories of \textit{crying baby}. On the other hand, DT reduces the false negatives, such as the \textit{knocking on door} event observed between 10 to 15 s.

\textbf{Limitation of PSELDNets.}
We observe the Fine-tune method exhibits an increase of more than $1^\circ$ in $\mathrm{LE}_\mathrm{CD}$ compared to the SOTA system of Shimada et al. \cite{shimada2021ensemble} on DCASE 2021 Task 3 and the SOTA system of Wang et al. \cite{wang2023dcase} on STARSS23, which can be attributed to the output temporal resolution of HTS-AT. Specifically, the output temporal resolution is 0.1 seconds in those systems \cite{wang2023dcase, shimada2021ensemble, hu2022track}, but approximately 0.3 seconds in HTS-AT due to the effect of split patches and the patch merging module. This discrepancy has minimal impact on the localization of static sources, for instance, in L3DAS22 Task 2, the Fine-tune method also shows performance improvement in $\mathrm{LE}_\mathrm{CD}$, compared to our previous system \cite{hu2022track}. However, it introduces a systematic error when localizing moving sources, such as those in DCASE 2021 Task 3 and STARSS23. On the other hand, PSELDNets provide more substantial performance improvement in detection than in localization. Moreover, the systems mentioned in this work \cite{shimada2021ensemble, nguyen2021dcase, hu2022track, wang2023four, xue2023dcase, wang2023dcase, hu22dw, wang2024dcase} have shown more significant improvement in detection than in localization. 

\subsection{Results of data-efficient fine-tuning}
\label{sec: res_deft}
In this section, we explore another application of PSELDNets: data-efficient fine-tuning. Specifically, we employ PSELDNets with AdapterBit in both low-resource-data and rich-source-data scenarios. Low-resource data refers to small synthetic datasets using only simulated SRIRs and no data augmentation technique, while rich-resource data includes substantial samples that have been augmented and derived from either real-world scenes or synthesis using collected SRIRs. All results are evaluated on the evaluation sets of DCASE 2021 Task 3 and L3DAS22 Task 2 and the validation set of STARSS23. 

Similar to Sec. \ref{sec: res_transfer_learning}, we also compare the \textit{From-scratch} method, the \textit{Fine-tune} method, and the additional \textit{AdapterBit} method. The Fine-tune method fine-tunes all parameters of PSELDNets, while the AdapterBit method only fine-tunes the inserted Adapter module and the bias item of PSELDNets.

\begin{table*}[t!]
    \centering
    \caption{Results of data-efficient fine-tuning on L3DAS22 Task 2.}
        \begin{tabular}{c|c|c|cccc|c}
            \toprule[1pt]
            Dataset & \# Channels & Method  & $\mathrm{ER}_{20^{\circ}}\downarrow$ & $\mathrm{F}_{20^{\circ}}\uparrow$ &
            $\mathrm{LE}_\mathrm{CD}\downarrow$ & $\mathrm{LR}_\mathrm{CD}\uparrow$ & $\mathcal{E}_{\mathtt{SELD}}\downarrow$\\
            \midrule 
             \multirow{6}{*}{\makecell{L3DAS22 - split\_ov1}} & 4 & From-scratch & 0.697& 30.1\%& $23.2^\circ$& 43.1\% & 0.523 \\
             & 4 & Fine-tune & 0.409& 64.3\%& $12.7^\circ$& 65.6\%& 0.295 \\
             & 4 & AdapterBit & 0.392& 64.7\%& $13.3^\circ$& 67.6\%& 0.286\\
             \cmidrule{2-8}
             & 1 & Fine-tune & 0.617& 32.7\% &$25.1^\circ$ & 60.4\% & 0.456 \\
             & 1 & LoRA & 0.605& 37.0\% &$24.7^\circ$ & 60.5\% & 0.442 \\
             & 1 & Adapter & 0.620& 34.4\% &$25.4^\circ$& 62.0\%& 0.449\\
             & 1 & AdapterBit & 0.591& 37.9\% &$24.5^\circ$& 62.0\%& 0.432\\
             \midrule
             \midrule
             \multirow{4}{*}{\makecell{Synthetic dataset}}& 4 & Fine-tune   & 0.603& 38.0\%& $23.6^\circ$& 56.4\% & 0.448\\
             & 4 & AdapterBit & 0.607 & 38.1\% & $23.5^\circ$ & 57.8\% &0.445 \\
             \cmidrule{2-8}
             & 1 & Fine-tune & 0.634& 35.1\%&$23.1^\circ$ & 49.0\% & 0.480\\
             & 1 & AdapterBit  & 0.618& 37.8\%&$22.1^\circ$& 51.7\%& 0.461 \\
             \bottomrule[1pt]
        \end{tabular}
    \label{tab: L3DAS22-mono}
\end{table*}

\subsubsection{Effect of AdapterBit}
Our ablation studies employ \textit{split\_ov1} subset of the L3DAS22 Task 2 dataset for training. This subset contains 250 30-second recordings without overlapping sound events. For training on monophonic clips, we extract the first-channel signal from FOA signals and then generate pseudo-FOA signals using these extracted monophonic signals. We evaluate the performance of four methods on pseudo-FOA signals, including Fine-tune, AdapterBit, Adapter (AdapterBit without bias-tuning), and LoRA \cite{lora}, as shown in the 1-channel part of the top block of Table \ref{tab: L3DAS22-mono}. Experimental results indicate the effectiveness of the designed Adapter, which achieves an $\mathcal{E}_{\mathtt{SELD}}$ of 0.449, compared to 0.456 for the Fine-tune method. Incorporating additional bias items into Adapter tuning leads to further performance improvement. Additionally, AdapterBit exhibits superior performance relative to LoRA.

Moreover, we compare the performance of pseudo-FOA signals with the corresponding FOA signals, denoted as channels of 4 in the top block of Table \ref{tab: L3DAS22-mono}. The primary difference between pseudo-FOA signals and FOA signals lies in inter-channel correlations. Our observations reveal that leveraging PSELDNets, all methods using only monophonic signals significantly outperform the From-scratch method using corresponding FOA signals, which only achieves $\mathcal{E}_{\mathtt{SELD}}$ of 0.523. When comparing the Fine-tune and AdapterBit methods using monophonic signals with the corresponding FOA signals, the primary performance difference lies in localization, due to similar performance in $\mathrm{LR}_\mathrm{CD}$ but a significant performance difference in $\mathrm{LE}_\mathrm{CD}$, such as $\mathrm{LE}_\mathrm{CD}$ of $12.7^\circ$ and $\mathrm{LR}_\mathrm{CD}$ of 65.6\% in the 4-ch Fine-tune method and $\mathrm{LE}_\mathrm{CD}$ of $25.1^\circ$ and $\mathrm{LR}_\mathrm{CD}$ of 60.4\% in the 1-ch Fine-tune method. Original FOA signals of the target scenario contain more information about the acoustic environment and microphone array than pseudo-FOA signals. 

\begin{table}[t]
    \centering
    \caption{Computational efficiency of various fine-tuning methods.}
        \begin{tabular}{c|cccc}
            \toprule[1pt]
                Method & \makecell{\# Tunable Params} & Inference MACs \\
                \midrule
                Fine-tune & 28.1 M & 5.76 G\\
                Adapter & 4.9 M & 6.67 G\\
                AdapterBit & 5.0 M & 6.67 G\\
                LoRA & 5.0 M & 5.76 G\\ 
            \bottomrule[1pt]
        \end{tabular}
    \label{tab: comput_effic}
\end{table}

Tables \ref{tab: L3DAS22-mono} and \ref{tab: comput_effic} present a comparison of the performance and computational efficiency of various fine-tuning methods, respectively. AdapterBit exhibits superior performance compared to LoRA and Adapter despite using the same scale of tunable parameters. However, it requires more inference multiply-accumulate operations (MACs) than Fine-tune and LoRA. Notably, LoRA achieves inference MACs comparable to Fine-tune, as its additional parameters can be merged into the original model weights. Techniques such as knowledge distillation \cite{gou2021knowledge}, model quantization \cite{polino2018model}, and model pruning \cite{tyagi2020second, zhang2022carrying} offer the potential to further reduce the parameter scale and computational overhead, but their investigation is beyond the scope of this study.

\subsubsection{Results on low-resource data}
Given the absence of polyphonic indications in each clip from STARSS23 and DCASE 2021 Task 3, we create scenario-specific datasets for training, as well as for L3DAS22 Task 2. Each dataset, consisting of 120 one-minute FOA audio clips, is generated using simulated SRIRs. The bottom block of Table \ref{tab: L3DAS22-mono}, Table \ref{tab: DCASE2021-mono}, and Table \ref{tab: STARSS23-mono} illustrate the results of data-efficient fine-tuning. Tests with either multi-channel or monophonic signals demonstrate the efficacy of AdapterBit relative to the Fine-tune method. Remarkably, AdapterBit tuning on monophonic signals reaches performance levels comparable to those from synthesized FOA signals, thus potentially simplifying the adaptation process to a target scene and diminishing the requirement for synthesizing multi-channel clips. This suggests that the performance difference between original FOA signals and pseudo-FOA signals, as shown in the top block of Table \ref{tab: L3DAS22-mono}, may indicate the discrepancy between the simulated environments and the acoustic conditions in the real world. Additionally, Tables \ref{tab: L3DAS22-mono}, \ref{tab: DCASE2021-mono} and \ref{tab: STARSS23-mono} show the capacity of PSELDNets to generalize across diverse microphone array setups through the FOA format without incorporating scenario-specific information, despite differences in microphone array setups between \textit{synthetic-training-set} and scenario-specific datasets.

\begin{table}[t]
    \centering
    \caption{Results of data-efficient fine-tuning on DCASE 2021 Task 3.}
    \begin{adjustbox}{width=\columnwidth,center}
        \begin{tabular}{c|c|cccc|c}
            \toprule[1pt]
            \# Channels & Method & $\mathrm{ER}_{20^{\circ}}\downarrow$ & $\mathrm{F}_{20^{\circ}}\uparrow$ &
            $\mathrm{LE}_\mathrm{CD}\downarrow$ & $\mathrm{LR}_\mathrm{CD}\uparrow$ & $\mathcal{E}_{\mathtt{SELD}}\downarrow$\\
            \midrule 
             4 & Fine-tune  &0.621& 40.4\%&$23.0^\circ$& 56.9\%& 0.444 \\
             4 & AdapterBit  &0.610 &43.4\% &$22.5^\circ$ & 61.5\% &0.422 \\
            \midrule
             1 & Fine-tune  & 0.594& 42.3\%&$22.5^\circ$& 58.7\%& 0.427 \\
             1 & AdapterBit & 0.595& 44.9\%&$21.5^\circ$& 59.2\%& 0.418 \\
             \bottomrule[1pt]
        \end{tabular}
    \end{adjustbox}
    \label{tab: DCASE2021-mono}
\end{table}

\begin{table}[t]
    \centering
    \caption{Results of data-efficient fine-tuning on the STARSS23 dataset.}
    \begin{adjustbox}{width=\columnwidth,center}
        \begin{tabular}{c|c|cccc|c}
            \toprule[1pt]
            \# Channels & Method & $\mathrm{ER}_{20^{\circ}}\downarrow$ & $\mathrm{F}_{20^{\circ}}\uparrow$ &
            $\mathrm{LE}_\mathrm{CD}\downarrow$ & $\mathrm{LR}_\mathrm{CD}\uparrow$ & $\mathcal{E}_{\mathtt{SELD}}\downarrow$\\
            \midrule 
             4 & Fine-tune & 0.732& 23.6\%&$22.4^\circ$& 41.4\%& 0.552 \\
             4 & AdapterBit  & 0.736& 23.5\%& $24.7^\circ$& 44.8\%& 0.548\\
            \midrule
             1 & Fine-tune & 0.719& 19.2\%&$26.5^\circ$& 37.3\% & 0.575 \\
             1 & AdapterBit & 0.745& 22.9\%&$25.9^\circ$& 43.3\% & 0.557 \\
             \bottomrule[1pt]
        \end{tabular}
    \end{adjustbox}
    \label{tab: STARSS23-mono}
\end{table}

\subsubsection{Results on rich-resource data}

\begin{figure}
    \begin{center}
        \includegraphics[width=\linewidth]{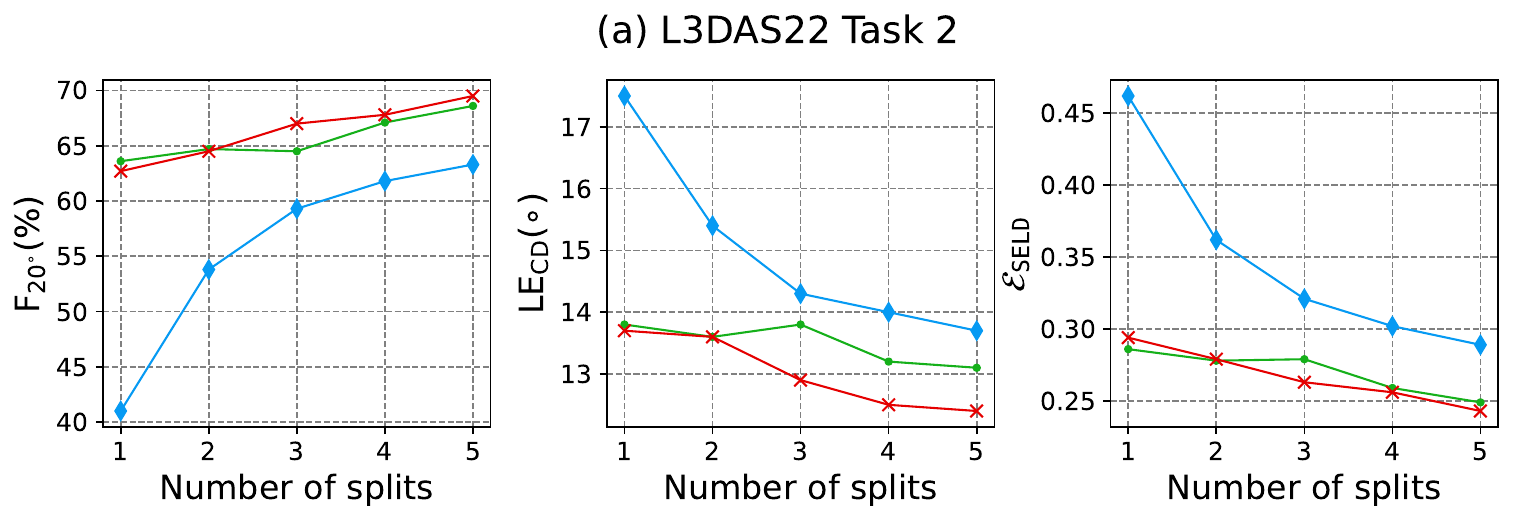}
    \end{center}
    \begin{center}
        \includegraphics[width=\linewidth]{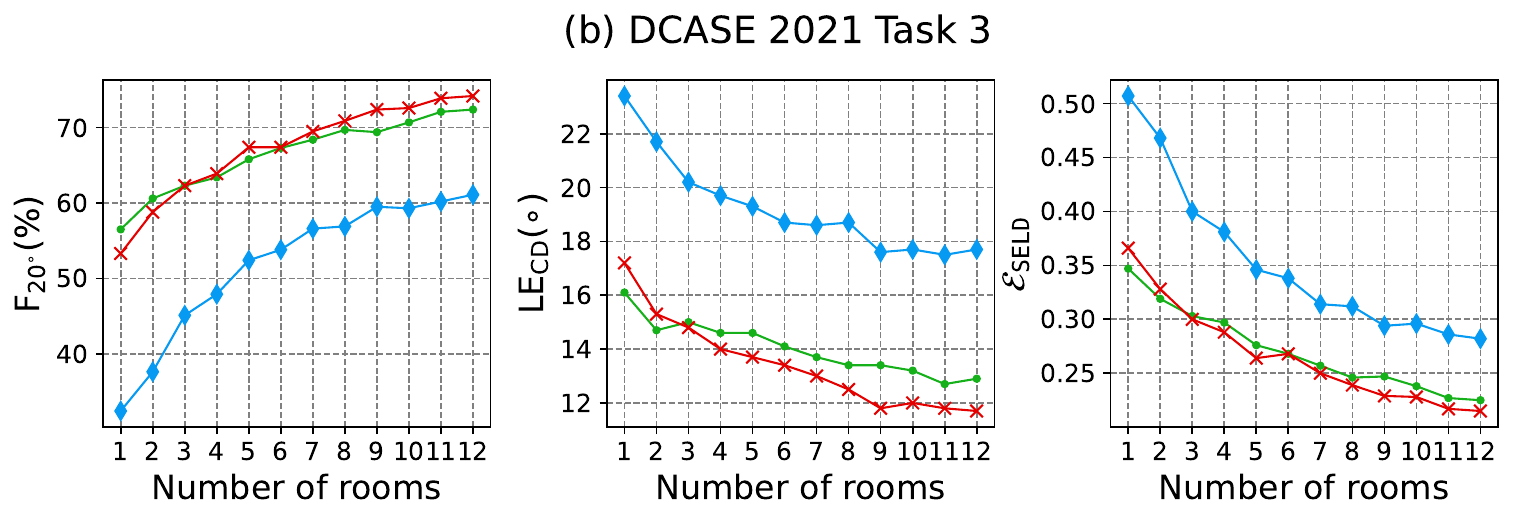}
    \end{center}
    \begin{center}
        \includegraphics[width=\linewidth]{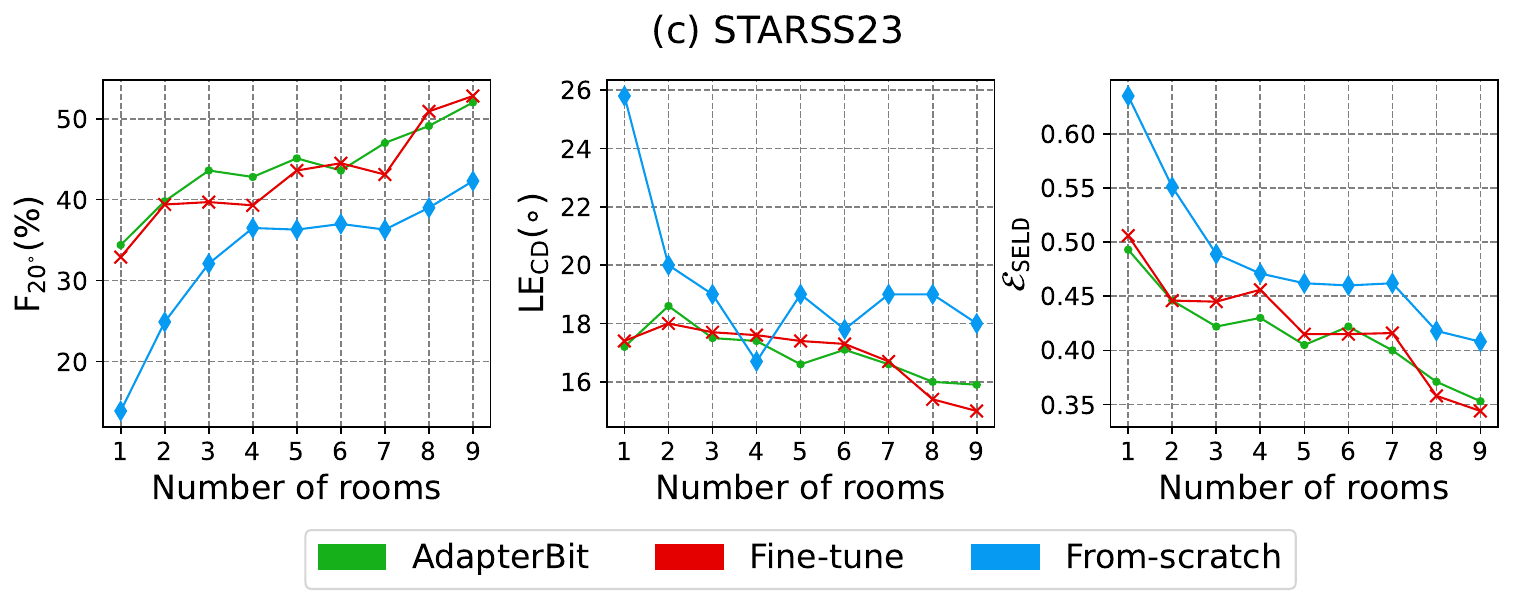}
    \end{center}
    \caption{The effect of AdapterBit on various proportions of the training set from downstream datasets.}
    \label{fig: multi_channel_adapterbit}
\end{figure}

Fig. \ref{fig: multi_channel_adapterbit} illustrates the performance of PSELDNets using AdapterBit, trained on varying proportions of the training set from downstream datasets, including clips from different numbers of rooms or splits. These datasets are either synthesized using measured SRIRs from real-world environments or directly collected in real-world environments. When employing STARSS23, a small-size synthesized dataset comprising spatial sound events of all corresponding classes is additionally utilized, because clips from any individual rooms in STARSS23 only encompass partial sound event classes. Experimental results indicate that both the Fine-tune method and AdapterBit exhibit superior performance compared to the From-scratch method, irrespective of the dataset size. Notably, AdapterBit also demonstrates more efficient data utilization than the Fine-tune method when trained on limited proportions of the training set, such as the clips from the first two splits of L3DAS22 Task 2, the first two rooms of DCASE 2021 Task 3 and the first five rooms of STARSS23. This aligns with the previous results on low-resource data. However, when trained on more data, AdapterBit's performance falls behind the Fine-tune method. This performance disparity can be attributed to the substantial distributional differences between the simulated environments and real-world acoustic environments, which AdapterBit may not effectively learn due to its limited trainable parameters.

\subsection{Results on Indoor Recordings}

\begin{figure}
    \begin{center}
        \subfloat{
        \includegraphics[width=0.47\linewidth]{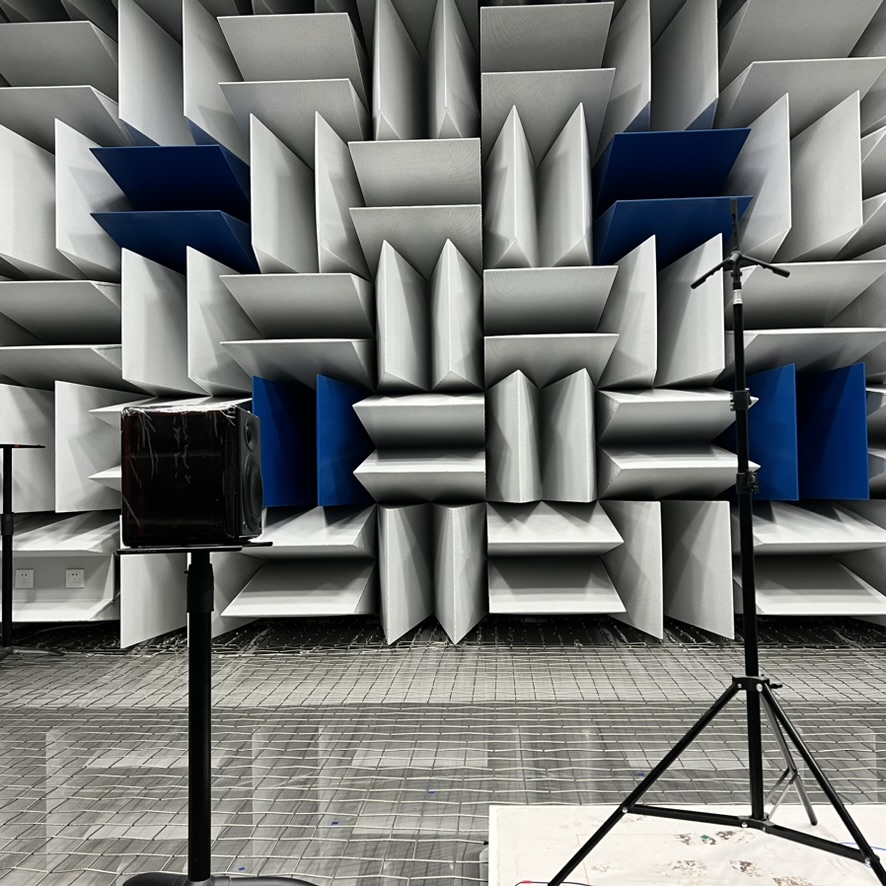}
        }
        \subfloat{
        \includegraphics[width=0.47\linewidth]{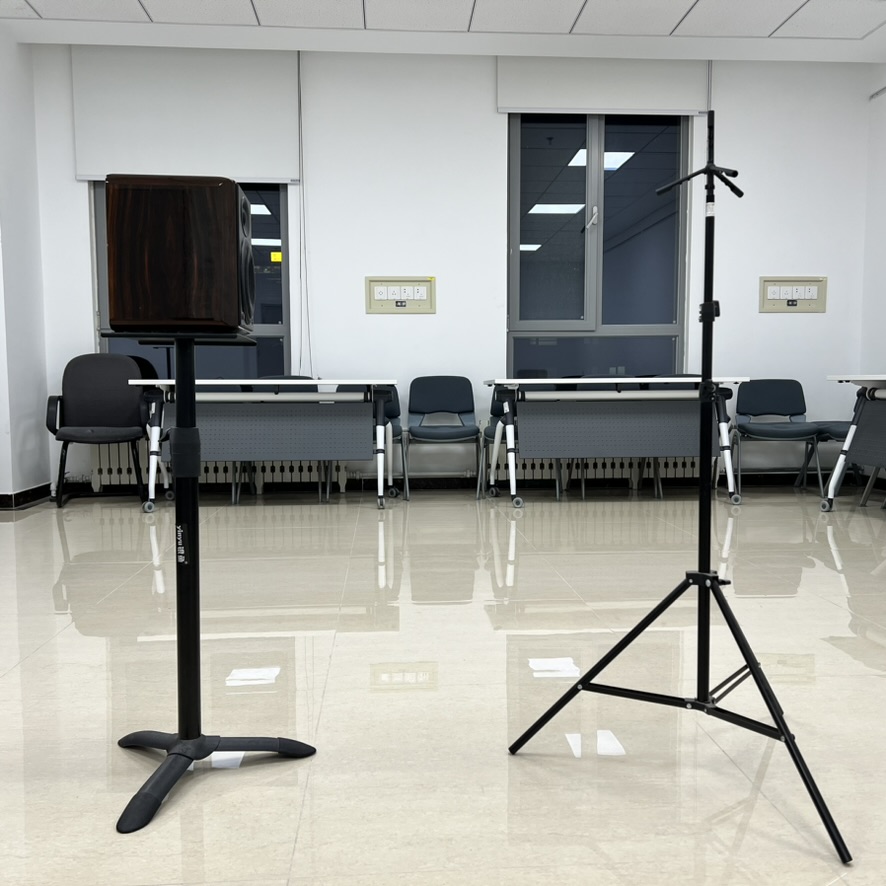}
        }
    \end{center}
    \caption{Recording environments used in real-world experiments, comprising an anechoic chamber and a meeting room.}
    \label{fig: recording_environments}
\end{figure}

In this section, we evaluate the transferability of PSELDNets and the impact of data-efficient fine-tuning on Indoor Recordings. As shown in Fig. \ref{fig: recording_environments}, we exploit a 4-channel unbaffled spherical microphone array with a radius of 0.12 m, arranged in a tetrahedral configuration, to record sound sources emitted by loudspeakers in two distinct environments: an anechoic chamber and a meeting room. In the meeting room, we estimate a reverberation time of $T_{60}=900$ ms and a signal-noise-ratio (SNR) of 6 dB. The microphone array was centrally positioned in a square, with loudspeakers placed at three vertical heights and eight predetermined horizontal locations corresponding to either the square's vertices or midpoints of the square's sides. The sides of the square have lengths of 4 m and 2.4 m. Therefore, there are a total of 48 sound source locations. We placed a loudspeaker among these locations and recorded one-minute audio clips with a maximum polyphony of 1 at each location, resulting in 48 non-overlapping audio clips. Additionally, we recorded 12 one-minute audio clips with a maximum polyphony of 2, where two loudspeakers were randomly placed among these 48 locations for each clip. We also measured the RIRs at 8 supplementary positions in the meeting room to facilitate efficient fine-tuning. The sound event clips from NIGENS \cite{nigens} are divided into training and test splits, with the training subset for data synthesis and the test subset for playback through the loudspeakers. In total, we collected 60 one-minute audio clips for each recording environment as the evaluation set.

We synthesize four datasets for training: Sim240, Sim120, Sim120\_ov1 and Co120. The maximum polyphony of Sim240, Sim120 and Co120 is 2, whereas the maximum polyphony of Sim120\_ov1 is 1. Sim240, Sim120 and Sim120\_ov1 are generated using simulated SRIRs and contain 240, 120, and 120 one-minute audio clips, respectively, while Co120 is synthesized using previously collected 8 RIRs in the meeting room and also contains 120 one-minute audio clips. The shape of the microphone array in synthetic datasets is consistent with the array used for recording in the real scene.

\begin{table*}[t]
    \centering
    \caption{Results on Indoor Recordings.}
    \begin{adjustbox}{width=\textwidth,center}
        \begin{tabular}{c|c|c|cccc|c|cccc|c}
            \toprule[1pt]
            \multirow{2}{*}{Method} & \multirow{2}{*}{\# Channels} & \multirow{2}{*}{Datasets} & \multicolumn{5}{c|}{Anechoic Chamber} & \multicolumn{5}{c}{Meeting Room} \\
             & & & $\mathrm{ER}_{20^{\circ}}\downarrow$ & $\mathrm{F}_{20^{\circ}}\uparrow$ &             $\mathrm{LE}_\mathrm{CD}\downarrow$ & $\mathrm{LR}_\mathrm{CD}\uparrow$ & $\mathcal{E}_{\mathtt{SELD}}\downarrow$ & $\mathrm{ER}_{20^{\circ}}\downarrow$ & $\mathrm{F}_{20^{\circ}}\uparrow$ &    $\mathrm{LE}_\mathrm{CD}\downarrow$ & $\mathrm{LR}_\mathrm{CD}\uparrow$ & $\mathcal{E}_{\mathtt{SELD}}\downarrow$\\
            \midrule
            From-scratch& 4& \makecell[c]{Sim240 + Sim120}& 0.392& 70.6\%& $12.9^\circ$& 73.9\%& 0.255& 0.448& 63.0\%& $15.5^\circ$& 71.2\%& 0.298\\
            Fine-tune& 4& \makecell[c]{Sim240 + Sim120}& 0.393& 69.3\%& $12.1^\circ$& 76.7\%& 0.250& 0.430& 64.4\%& $15.3^\circ$& 76.4\%& 0.277\\
            Fine-tune& 4& \makecell[c]{Sim240 + Co120}& 0.358& 72.8\%& $11.9^\circ$& 79.9\%& 0.224& 0.414& 67.6\%& $14.0^\circ$& 79.7\%& 0.255\\
            \midrule
            \midrule
            Fine-tune& 4& Sim120\_ov1& 0.526& 57.9\%& $13.9^\circ$& 65.2\%& 0.343& 0.686& 36.7\%& $24.7^\circ$& 67.3\%& 0.446\\
            AdapterBit & 4 & Sim120\_ov1& 0.480& 60.7\%& $14.1^\circ$& 63.2\%& 0.330& 0.615& 41.1\%& $23.5^\circ$& 65.3\% &0.420\\
            \midrule
            Fine-tune& 1& Sim120\_ov1& 0.626& 43.5\%& $22.0^\circ$& 53.3\%& 0.445& 0.809& 16.9\%& $39.7^\circ$& 50.7\%& 0.588\\
            AdapterBit& 1& Sim120\_ov1& 0.570& 48.4\%& $22.2^\circ$& 64.3\%& 0.392& 0.768& 22.2\%& $35.1^\circ$& 59.5\%& 0.537\\
            \bottomrule[1pt]
        \end{tabular}
    \end{adjustbox}
    \label{tab: recordings}
\end{table*}

Same as Sec. \ref{sec: res_deft}, we compare the \textit{Fine-tune}, \textit{From-scratch}, and \textit{AdapterBit} methods. The top block of Table \ref{tab: recordings} presents the transferability of PSELDNets on Indoor Recordings. We fine-tune PSELDNets on augmented synthetic datasets and evaluate the performance on Indoor Recordings collected from two environments. Notably, the distribution differences in sound event clips and recording locations between these two environments are minimal, excluding acoustic properties, such as noise level and reverberation. We observe that the performance in $\mathrm{LR}_\mathrm{CD}$ between the two environments is similar, but the performance difference is mainly in localization, which can be due to differences in acoustic environments. Experimental results demonstrate the effectiveness of PSELDNets. Additionally, replacing Sim120 with Co120 can improve performance in both localization and detection, since collected RIRs carry more information about the acoustic environment and microphone characteristics. On the other hand, we note that $\mathrm{LE}_\mathrm{CD}$ in the ideal acoustic environment, anechoic chamber, reaches approximately $12^\circ$, perhaps due to the large radius and unbaffled configuration of the spherical microphone array \cite{rafaely2015fundamentals}, which leads to incorrect FOA conversion, especially in the high-frequency range. Furthermore, while we transfer pre-trained models to specific scenarios, the considerable variation in distribution between the source domain and the target domain \cite{survey-transfer-learning}, such as the microphone characteristics, puts the prior knowledge at high risk of losing effectiveness \cite{grumiaux2022survey, survey-transfer-learning, DA-survey, DA-survey-tpami}. 

The bottom block of Table \ref{tab: recordings} illustrates the efficacy of the data-efficient fine-tuning method on Indoor Recordings. The results indicate that the model fine-tuned on FOA signals performs better than on pseudo-FOA signals by a large margin, which can be attributed to substantial differences in microphone array configurations and acoustic environments between \textit{synthetic-training-set} and the synthetic and recorded datasets used in this section. Additionally, when using multi-channel or monophonic signals, AdapterBit tuning performs better than the Fine-tune method. Notably, it suggests that AdapterBit is particularly effective when only using monophonic signals. We hypothesize that AdapterBit prevents catastrophic interference \cite{adaptformer}, thereby decreasing the likelihood that the model forgets previously learned knowledge when adapting to new scenarios.

\section{Conclusion}
This paper has built sound event localization and detection (SELD) foundation models by introducing pre-trained SELD networks (PSELDNets) on a large-scale synthetic dataset. The synthetic dataset encompasses 1,167 hours of audio recordings with an ontology of 170 sound classes. To enhance the adaptability of PSELDNets to specific scenarios with low-resource data, we have presented AdapterBit, a data-efficient fine-tuning technique. We evaluate PSELDNets on \textit{synthetic-test-set} and achieve satisfactory performance. We transfer PSELDNets to several downstream datasets that are publicly available, as well as to our own collected recordings, Indoor Recordings. The experimental results demonstrate superior performance compared to previous state-of-the-art systems. Moreover, incorporating AdapterBit into PSELDNets enhances the efficiency of the transferability for low-resource data, including both limited multi-channel and monophonic audio clips. Future work will focus on model compression and computational efficiency to enable real-time applications.

\scriptsize
\bibliographystyle{IEEEtran}
\bibliography{refs} 

\begin{IEEEbiography}[{\includegraphics[width=1in,height=1.25in,clip,keepaspectratio]{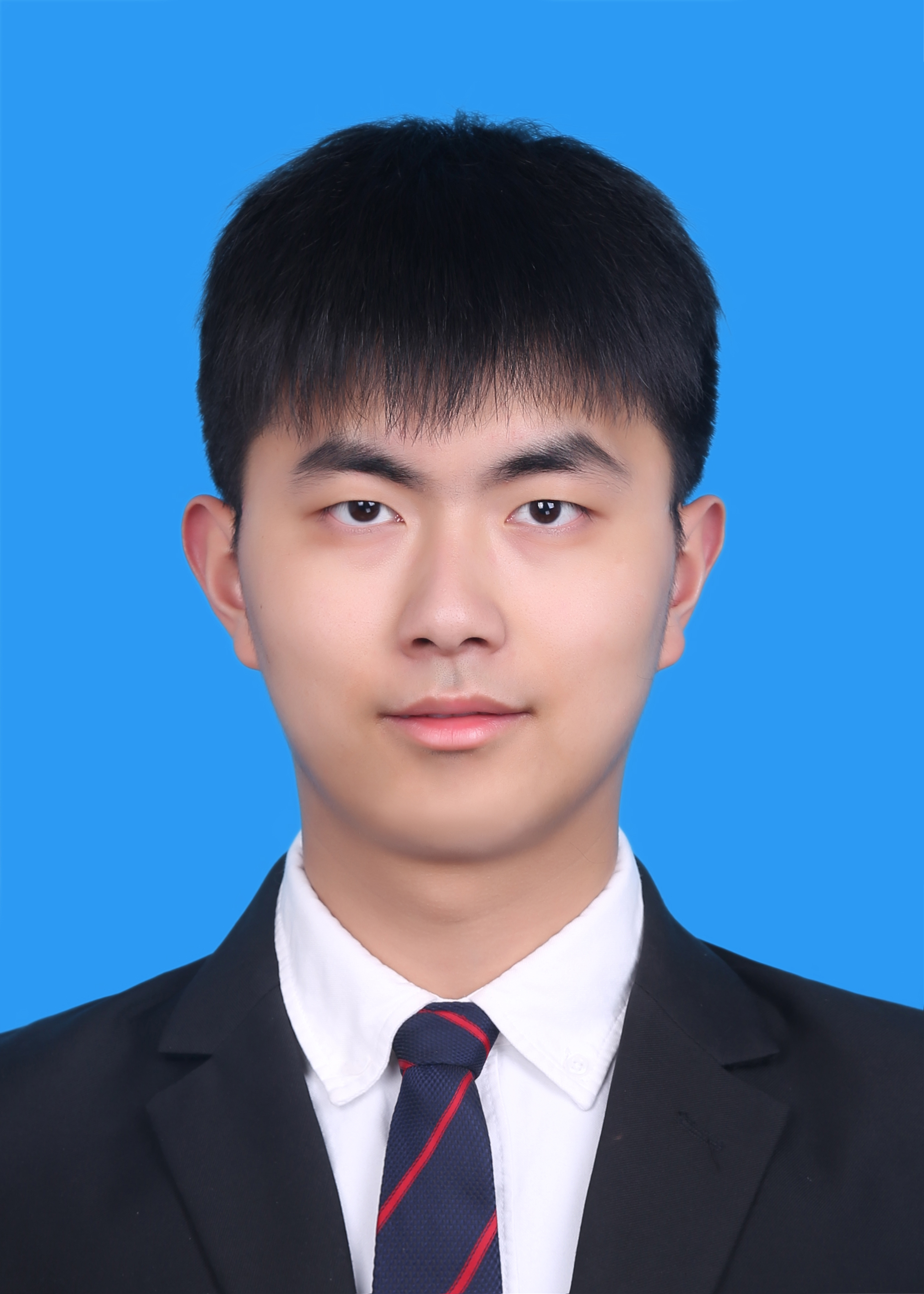}}]{Jinbo Hu}
(S'22) received the B.E. degree from Hubei University, Wuhan, China, in 2020. He is currently working towards the Ph.D. degree with the Institute of Acoustics, Chinese Academy of Sciences, Beijing, China. His research interests include sound event localization and detection, audio signal processing and deep learning. He was the winner of ICASSP 2022 Grand Challenge L3DAS22 Task 2: 3D Sound Event Localization and Detection and obtained the second place in the team ranking in DCASE 2022 Challenge Task 3: Sound Event Localization and Detection Evaluated in Real Spatial Sound Scenes.
\end{IEEEbiography}

\vskip 0pt plus -1fil

\begin{IEEEbiography}[{\includegraphics[width=1in,height=1.25in,clip,keepaspectratio]{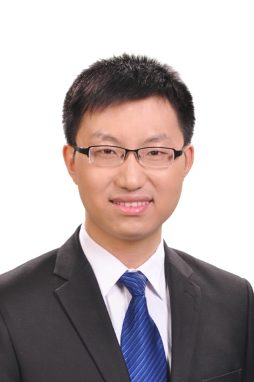}}]{Yin Cao}
(M'18) is currently an associate professor at the Department of Intelligent Science at Xi'an Jiaotong-Liverpool University. He received a B.Sc. degree in Electronic Science and Engineering from Nanjing University, China, in 2008 and a Ph.D. degree from the Institute of Acoustics, Chinese Academy of Sciences, China, in 2013. He then worked in the Acoustics group at Brigham Young University, US, and at the Institute of Acoustics, Chinese Academy of Sciences, China. In 2018, he joined Centre for Vision, Speech, and Signal Processing at the University of Surrey. His research topics include machine learning and signal processing for audio, speech, and acoustics. He was the winner of urban sound tagging in detection and classification of acoustic scenes and events (DCASE) 2020 challenge and achieved second-best of sound event detection and localization tasks in DCASE 2019 challenge. He also won several DCASE challenges and ICASSP challenges. He has served as an Associate Editor for Noise Control Engineering Journal since 2020. He is a frequent reviewer for IEEE/ACM Transactions on Audio, Speech, and Language Processing.
\end{IEEEbiography}

\vskip 0pt plus -1fil

\begin{IEEEbiography}[{\includegraphics[width=1in,height=1.25in,clip,keepaspectratio]{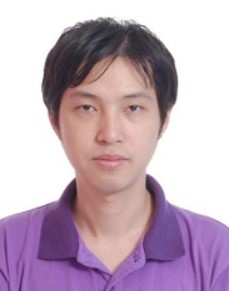}}]{Ming Wu}
(M'13) received the B.E. degree in electronics and the Ph.D. with a dissertation on active noise control from Nanjing University, Nanjing, China, in 2002 and 2007, respectively. From 2007 to 2008, he was a Postdoctoral Researcher with the Institute of Acoustics, Nanjing University. Since 2008, he has been with the Institute of Acoustics, Chinese Academy of Sciences, Beijing, China, as an Associate Professor. In 2016, he became a professor. He is currently the Deputy Director of the Key Laboratory of Noise and Vibration Research, Institute of Acoustics, Chinese Academy of Sciences. His main research interests include noise and vibration control, sound field perception and control, electroacoustics, and audio signal processing.
\end{IEEEbiography}

\vskip 0pt plus -1fil

\begin{IEEEbiography}[{\includegraphics[width=1in,height=1.25in,clip,keepaspectratio]{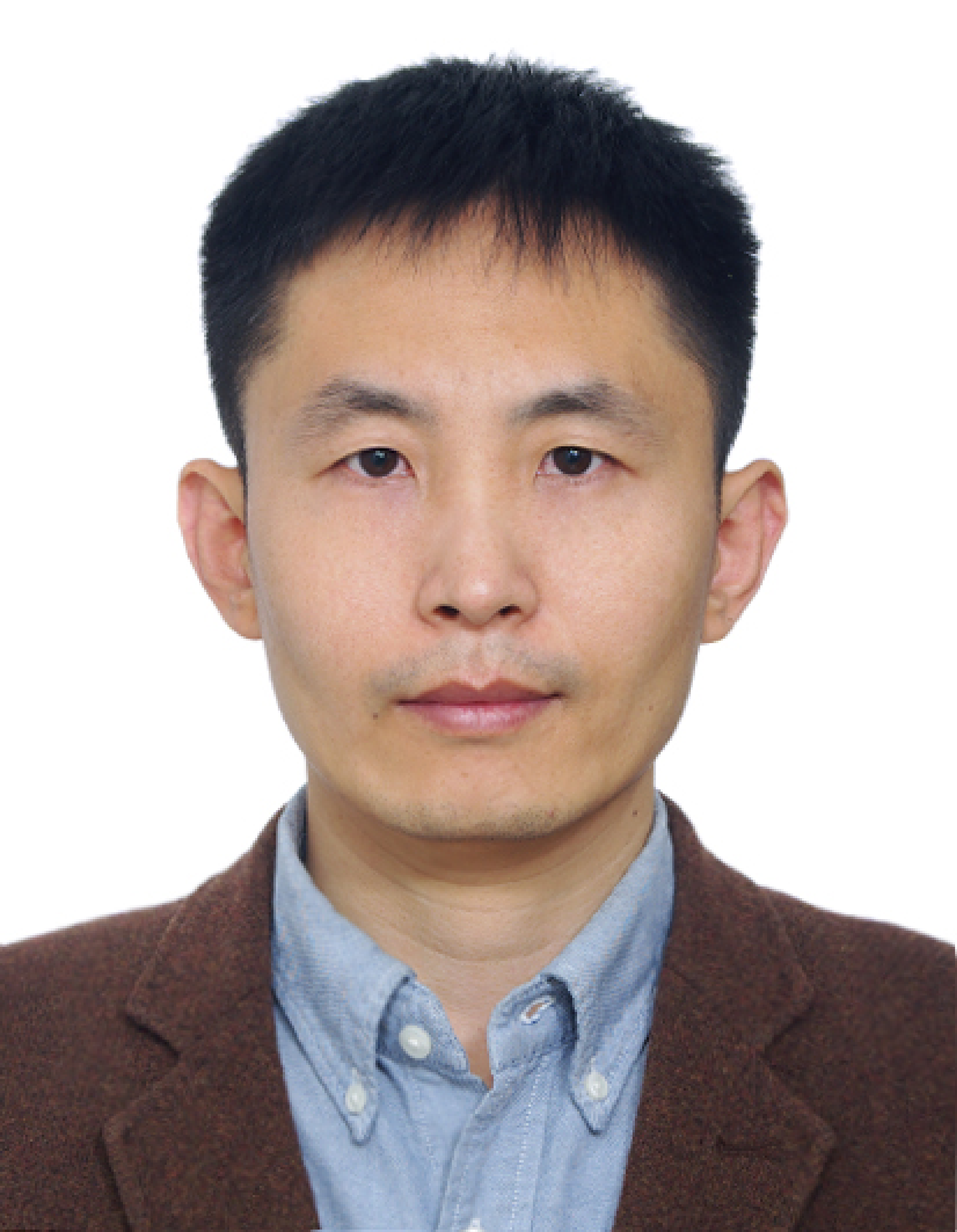}}]{Feiran Yang}
(M'14) received the B.E. degree in electrical engineering from Shandong University, Jinan, China, in 2005, the M.E. degree in signal processing from Southeast University, Nanjing, China, in 2008, and the Ph.D. degree in signal processing from the Institute of Acoustics, Chinese Academy of Sciences (IACAS), Beijing, China, in 2013. From March 2008 to June 2010, He was with Fortemedia, Inc., as a DSP engineer. From April 2016 to March 2017, he was with Ruhr-Universit\"at Bochum as a Visiting Scholar. Since 2013, he has been with IACAS, where he is currently a Professor. His research interests are adaptive filtering, microphone array signal processing, and spatial audio. He received the President's Award of the Chinese Academy of Sciences in 2013. He also received the Excellent Doctoral Dissertation Award of the Chinese Academy of Sciences in 2016.
\end{IEEEbiography}

\vskip 0pt plus -1fil

\begin{IEEEbiography}[{\includegraphics[width=1in,height=1.25in,clip,keepaspectratio]{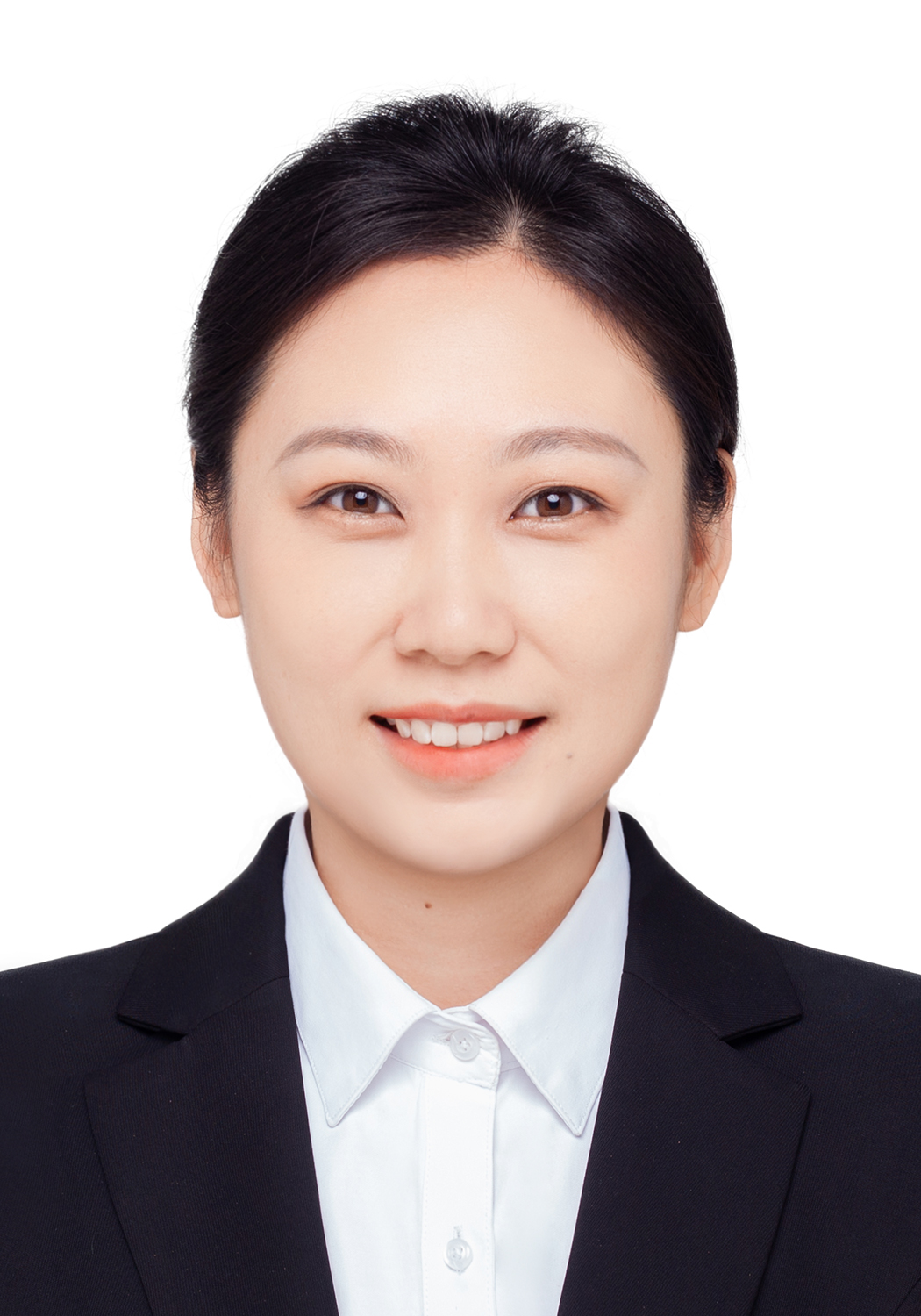}}]{Fang Kang}
received the B.E. degree in Communication Engineering from China University of Geosciences, Wuhan, China, in 2016, and the Ph.D. degree in Signal and Information Processing from the Institute of Acoustics, University of Chinese Academy of Sciences, Beijing, China, in 2021. From 2021 to 2023, she was an Audio Algorithm Researcher with Tencent, China. From 2023 to 2024, she was an Assistant Professor with the School of Advanced Technology, Xi’an Jiaotong-Liverpool University, Suzhou, China. Since 2024, she has been a Postdoctoral Researcher with the Center for Machine Vision and Signal Analysis (CMVS), University of Oulu, Oulu, Finland. Her research interests include speech and audio signal processing, blind source separation, sound event detection, and multimodal generation. She organized the 3rd MiGA-IJCAI Challenge and served as a co-chair of the 3rd MiGA Workshop at IJCAI 2025.
\end{IEEEbiography}

\vskip 0pt plus -1fil

\begin{IEEEbiography}[{\includegraphics[width=1in,height=1.25in,clip,keepaspectratio]{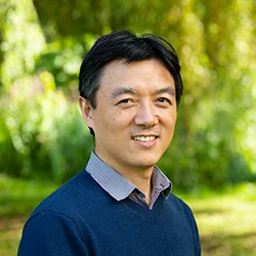}}]{Wenwu Wang}
(M’02-SM’11) was born in Anhui, China. He received the B.Sc., M.E., and the Ph.D. degrees, all in the field of automation, from Harbin Engineering University, China, in 1997, 2000, and 2002, respectively. He then worked with King’s College London, Cardiff University, Tao Group Ltd. (now Antix Labs Ltd.), and Creative Labs, before joining University of Surrey, U.K., in May 2007, where he is currently a Professor in Signal Processing and Machine Learning, and an Associate Head in External Engagement, School of Computer Science and Electronic Engineering, University of Surrey, UK. He is also an AI Fellow at the Surrey Institute for People Centred Artificial Intelligence. His current research interests include signal processing, machine learning and perception, artificial intelligence, machine audition (listening), human-AI collaboration, and statistical anomaly detection. He has (co)-authored over 400 papers in these areas. His works have been recognized with various awards, including the 2022 IEEE Signal Processing Society Young Author Best Paper Award, ICAUS 2021 Best Paper Award, DCASE 2020, 2023 and 2024 Judge’s Award, DCASE 2019 and 2020 Reproducible System Award, and LVA/ICA 2018 Best Student Paper Award. He is an Associate Editor (2024-2026) for IEEE Transactions on Multimedia. He was a Senior Area Editor (2019-2023) and Associate Editor (2014-2018) for IEEE Transactions on Signal Processing, and an Associate Editor (2020-2025) for IEEE/ACM Transactions on Audio Speech and Language Processing. He was the elected Chair (2023-2024) of IEEE Signal Processing Society (SPS) Machine Learning for Signal Processing (MLSP) Technical Committee, and a Board Member (2023-2024) of IEEE SPS Technical Directions Board. He is currently the elected Chair (2025-2027) of the EURASIP Technical Area Committee on Acoustic Speech and Music Signal Processing, and an elected Member (2021-2026) of the IEEE SPS Signal Processing Theory and Methods Technical Committee. He was on the organization committee of IEEE ICASSP 2019 and 2024, INTERSPEECH 2022, IEEE MLSP 2013 and 2024, and IEEE SSP 2009. He is a Technical Program Co-Chair of IEEE MLSP 2025. He has been a keynote or plenary speaker at more than 20 international conferences and workshops.
\end{IEEEbiography}

\vskip 0pt plus -1fil

\begin{IEEEbiography}[{\includegraphics[width=1in,height=1.25in,clip,keepaspectratio]{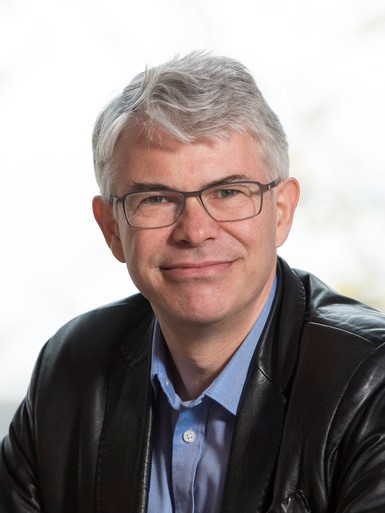}}]{Mark D. Plumbley}
(S'88-M'90-SM'12-F'15) received the B.A.(Hons.) degree in electrical sciences and the Ph.D. degree in neural networks from University of Cambridge, Cambridge, U.K., in 1984 and 1991, respectively. He is Professor of Signal Processing at the Centre for Vision, Speech, and Signal Processing at the University of Surrey, Guildford, U.K. His current research concerns AI, machine learning and signal processing for analysis, recognition and generation of sound. He led the first international data challenge on Detection and Classification of Acoustic Scenes and Events, and currently holds an Engineering and Physical Sciences Research Council Fellowship on ``AI for Sound''. He is a Member of the IEEE Signal Processing Society Technical Committee on Audio and Acoustic Signal Processing, and a Fellow of the IET and IEEE.
\end{IEEEbiography}

\vskip 0pt plus -1fil

\begin{IEEEbiography}[{\includegraphics[width=1in,height=1.25in,clip,keepaspectratio]{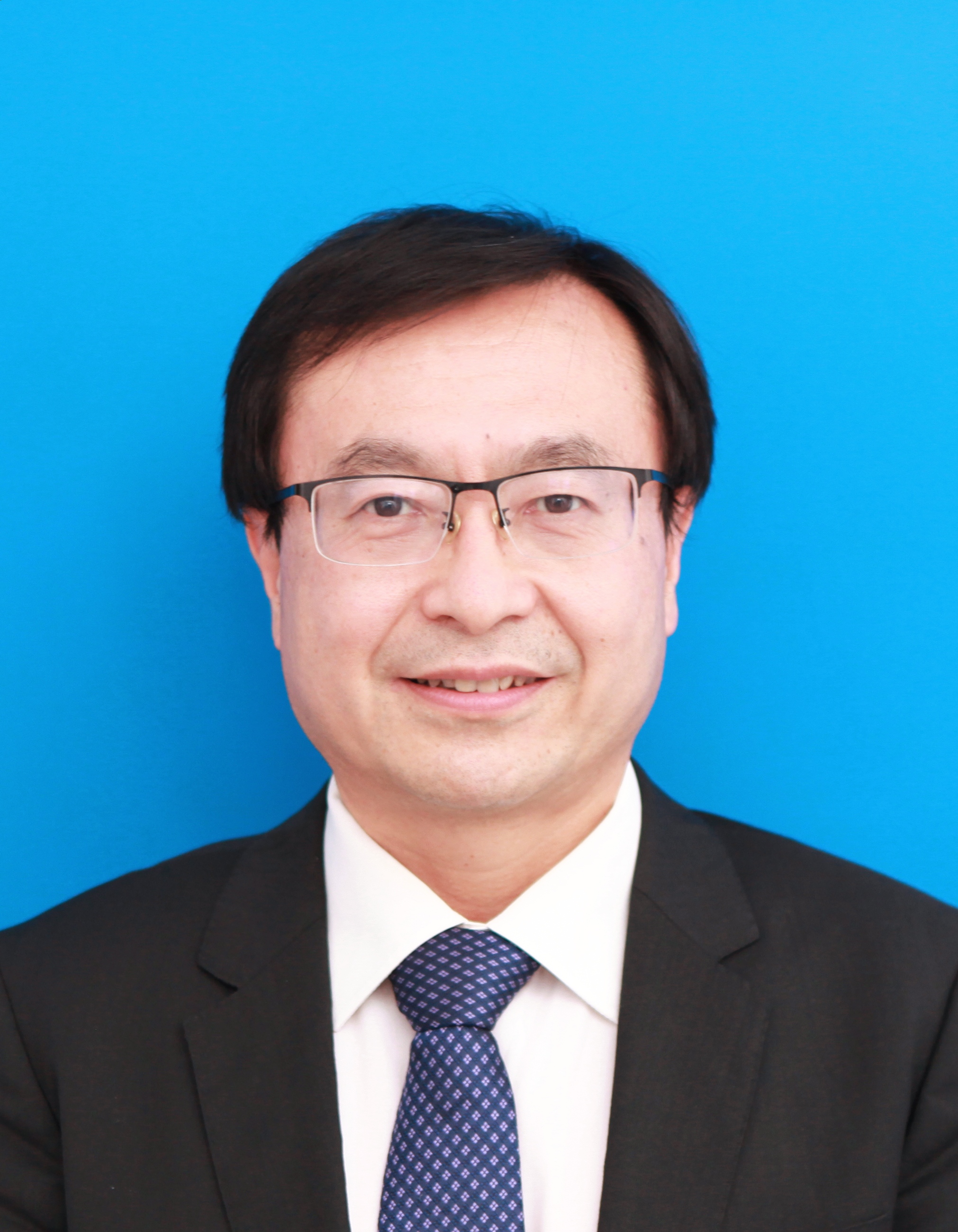}}]{Jun Yang}
(M'99-SM'04) received the B.Eng. and M. Eng. degrees from Harbin Engineering University, Harbin, China, and the Ph.D. degree in acoustics from Nanjing University, Nanjing, China, in 1990, 1993, and 1996, respectively. From 1996 to 1998, he was a Postdoctoral Fellow with the Institute of Acoustics, Chinese Academy of Sciences (IACAS), Beijing, China. From October 1998 to April 1999, he was with Hong Kong Polytechnic University as a Visiting Scholar. From 1997 to 1999, he was with IACAS as an Associate Professor. He joined the School of Electrical and Electronic Engineering, Nanyang Technological University, Singapore, as a Research Fellow, a Teaching Fellow, Assistant Professor, and Associate Professor in 1999, 2001, 2003, and 2005, respectively. Since November 2003, he has been a Professor at IACAS. From 2011 to 2020, he was the Director of the Key Laboratory of Noise and Vibration Research, Institute of Acoustics, Chinese Academy of Sciences. He is currently the Deputy Director of IACAS and Associate Dean of School of Electronic, Electrical and Communication Engineering, University of Chinese Academy of Sciences, China. His main research interests include communication acoustics, 3-D audio systems, acoustic signal processing, sound field control, and nonlinear acoustics. He is a Fellow of the International Institute of Acoustics and Vibration, Acoustic Society of China (ASC) and Chinese Institute of Electronics. He is the Vice President of the International Institute of Noise Control Engineering, Vice President of and Secretary-General of ASC, and serves as the Editor-in-Chief of Sound \& Vibration Journal.  
\end{IEEEbiography}

\vskip 0pt plus -1fil

\end{document}